\pdfoutput=1
\documentclass[preprint,12pt,times]{elsarticle}
\usepackage{amssymb}
\usepackage{amsmath}
\usepackage{graphics}
\usepackage{graphicx}
\usepackage{subfigure}
\usepackage{dcolumn}
\usepackage{bm}
\usepackage{hyperref}
\usepackage{color}

\journal{Advances in Water Resources}

\begin{document}

\begin{frontmatter}

\title{Chaotic Advection at the Pore Scale: Mechanisms, Upscaling and Implications for Macroscopic Transport}

\author[RMIT]{D.R. Lester\fnref{label2}}
\ead{daniel.lester@rmit.edu.au}
\fntext[label2]{ph: +61 3 9925 2404}
\author[CLW,CMR,UWA]{M.G. Trefry}
\author[SWIN,CMAN,MON]{Guy Metcalfe}
\address[RMIT]{Department of Chemical Engineering, School of Engineering, Royal Melbourne Institute of Technology, Melbourne, Victoria 3001, Australia}
\address[CLW]{CSIRO Land and Water, Floreat, Western Australia 6014, Australia}
\address[CMR]{CSIRO Mineral Resources, Floreat, Western Australia 6014, Australia}
\address[UWA]{School of Mathematics and Statistics, University of Western Australia, Perth, Western Australia 6009, Australia}
\address[SWIN]{Department of Mechanical and Product Design
  Engineering, Swinburne University of Technology, Hawthorn, Victoria
  3122, Australia}
\address[CMAN]{CSIRO Manufacturing, Clayton, Victoria 3169, Australia}
\address[MON]{School of Mathematical Sciences, Monash University,
  Clayton, Victoria 3800, Australia}

\begin{abstract}
  The macroscopic spreading and mixing of solute plumes in saturated
  porous media is ultimately controlled by processes operating at the
  pore scale.  Whilst the conventional picture of pore-scale
  mechanical dispersion and molecular diffusion leading to persistent
  hydrodynamic dispersion is well accepted, this paradigm is
  inherently two-dimensional (2D) in nature and neglects important
  three-dimensional (3D) phenomena.  We discuss how the kinematics of
  steady 3D flow at the pore-scale generate \emph{chaotic
    advection}---involving exponential stretching and folding of fluid
  elements---the mechanisms by which it arises and implications of
  microscopic chaos for macroscopic dispersion and mixing.  Prohibited
  in steady 2D flow due to topological constraints, these phenomena
  are ubiquitous due to the topological complexity inherent to all 3D
  porous media.  Consequently 3D porous media flows generate
  profoundly different fluid deformation and mixing processes to those
  of 2D flow. The interplay of chaotic advection and broad transit
  time distributions can be incorporated into a continuous-time random
  walk (CTRW) framework to predict macroscopic solute mixing and
  spreading.  We show how these results may be generalised to real
  porous architectures via a CTRW model of fluid deformation, leading
  to stochastic models of macroscopic dispersion and mixing which both
  honour the pore-scale kinematics and are directly conditioned on the
  pore-scale architecture.
\end{abstract}

\begin{keyword}
poroous media, dispersion, dilution, mixing, upscaling, chaotic advection
\end{keyword}

\end{frontmatter}

\section{Introduction}
The migration and dispersion of solutes in groundwater systems is
fundamental to chemical and biological reaction in the
regolith. Solute transport in porous media is a well studied topic,
and relevant dispersion, dilution and mixing process models have
utility across a range of transport and reaction scenarios, including
in contaminant remediation, in extraction of commodities (minerals,
oil and gas), in surface water - groundwater interaction, and so on.
In fact, more generally, fluid dispersion, dilution and mixing belong
to a broad class of \textit{dissipative processes} that mediate
structure and instability in the geosciences, where thermal,
hydrological, mechanical and chemical (THMC) phenomena are coupled
over wide ranges of spatial and temporal scales~\cite{RegenauerLieb:2013aa,RegenauerLieb:2013ab}. Evidence is
growing that fluid dissipation processes manifesting at one scale can
have profound influences on mechanical, thermal and chemical
structures at larger scales. It is important to consider whether our
understandings of solute transport in porous media are sufficient to
underpin the rich set of dependent phenomena.

If we focus on the problem of solute transport in groundwater systems,
the last forty years have seen growing sophistication of conceptual
and theoretical tools, observations and predictions of spreading and
reaction in solute plumes. Even so, there is still much to learn about
how fluids and solutes migrate and mix in porous media, and how we may
control or manipulate these transport processes for benefit.  Despite
recent advances in dynamical engineering of subsurface reactions at
the macroscale \cite{Trefry:2012aa, Mays:2012}, significant questions
about the scale dependence of plume dispersion, about solute mixing
and about \textit{in situ} biogeochemical reaction kinetics remain
unresolved.  It is our thesis that the resolution of these
meso/macroscale questions lies at the pore scale, specifically in
terms of the complex interplay between the pore space and fluid
mechanics, and how this interplay upscales into observed plume
characteristics. We contend that pore-scale dynamics engender
characteristics of transport and mixing that persist through upscaling
into the macroscopic domain and therefore play an important role in
conditioning effective mixing and reaction phenomena in porous media.

Computer tomography (CT) has enabled significant advances in the understanding and prediction of pore scale fluid percolation, transport and reaction phenomena~\cite{Blunt:2013aa}. Using three-dimensional (3D) micro-CT scanning, it is possible to construct digital representations of pore spaces from physical samples, and then use numerical flow and transport solvers to track and predict fluid-borne phenomena in three dimensions. For example, Liu and Regenauer-Lieb \cite{Liu:2011} show how digital cores can be post-processed to characterise percolation thresholds and critical indices, which permits detailed upscaling of microscopic sample observations to effective macroscopic material properties. Kang et al.~\cite{Kang:2015} use low-Reynolds number Navier-Stokes computations to show how the pore fluid velocity bursts observed in strongly heterogeneous systems are correlated with anomalous transport signatures at the micron scale in real rock samples. In a larger scale study, Alhashmi et al. \cite{Alhashmi:2015} have imaged spherical bead assemblies and generated Stokes flow fields in the interstitial spaces. From these fields they were able to reconcile under-prediction of effective bimolecular reaction rates in terms of incomplete fluid mixing in the complex pore space. The digital imaging approach is also being used to progress detailed simulations of multiphase processes in complex pore geometries at the micron scale \cite{Raeini:2015}. Together, these and other studies have shown that it is possible to link the microscale to the macroscale through a combination of high-resolution imaging, direct computation and upscaling techniques. However, dealing simultaneously with the complexities of topology, geometry and fluid reactions may obscure some of the component processes. 

Stochastic models offer a means to deal with both the uncertainty and complexity of the both the pore-scale architecture and macroscopic heterogeneities, and to facilitate upscaling. The kernels of the stochastic models embody the physical properties of both the porous medium and the hosted processes. Whilst these models possess a great deal of flexibility with respect to capturing a wide range of (often anomalous) transport, mixing and reaction phenomena, it is important to develop strong linkages between the physical processes and the model kernels. The alternative approach is to create stochastic models with ``black box'' kernels derived from regression analysis, which provide limited physical insight into the process at hand, and hamper development of generalised models which may be readily applied to different media or chemical kinetics. This shortcoming often manifests as an upscaling failure - i.e. models which are conditioned upon observations at a particular length-scale fail at others, or cannot be readily translated to regimes with different Pecl\'{e}t or Damk\"{o}hler numbers. Central to the development of so-called \emph{physically conditioned} stochastic models is the need to derive models which capture the underlying fundamental dynamics and so recover the associated constraints and physical phenomena. As the kernels of these models are more tightly constrained they exhibit a reduced risk of spurious regression, are directly linked to the underlying physical processes, and so are more readily generalised across scales, media and processes.

In this paper we consider how the kinematics of steady 3D flow at the
pore-scale influence macroscopic fluid mixing and chemical transport
in saturated porous media, and how to develop upscaled models of
dispersion and dilution which capture these processes. Our purpose
here is to strengthen the conceptual linkages between pore-scale
physics and macroscopic observations of solute transport and mixing in
porous media. This paper describes concepts of pore scale topology and
how they relate to transport of solutes in porous
media. Dimensionality is a critical parameter. In contrast with common
conceptual pictures of two dimensional branching of fluid streamlines
used as a motivation for hydrodynamic dispersion, it is shown that
proper consideration of truly 3D flow and transport at the pore scale
admits wholly new transport mechanisms which are not readily apparent
in conventional dispersion models. Specifically, 3D steady flow opens
the door to the possibility of chaotic fluid trajectories
\cite{Ottino:1989aa}, whereby fluid elements deform exponentially in
time, leading to wholly different mixing and dispersion. Indeed,
chaotic advection has been shown \cite{Lester:2013aa} to be inherent
to steady 3D flow at the pore scale.  Conversely, these dynamics are
inadmissible in steady 2D flow, and so are neglected---or in fact not
suspected to be needed---in conventional modelling frameworks.

Such chaotic (space-filling) transport dynamics are quite distinct from the conventional picture of hydrodynamic dispersion and can lead to persistent features that manifest in solute plumes at much greater spatial and temporal scales. The paper provides examples of persistent pore-scale chaotic phenomena, and then discusses some implications of pore-scale chaos for macroscopic transport. This topological approach to fluid physics at the pore scale is a relatively new area and opportunities abound for further development. The remainder of this paper is organised as follows; Section~\ref{sec:porescale_flow_deform} discusses topology of the pore pace and implications for flow, Section~\ref{sec:pore_chaos} discusses the mechanisms which lead to chaotic advection at the pore scale, and in Section~\ref{sec:network_model} fluid flow and transport are considered in a model open porous network. In Section~\ref{sec:pore_mixing} pore-scale mixing due to the interplay of chaotic advection and molecular diffusion is quantified, and the impact of these processes upon longitudinal dispersion is considered in Section~\ref{sec:long_disperse}. A discussion of approaches to extend the derived results to real porous media is presented in Section~\ref{sec:real_media}, and the paper concludes with some commentary on particularly promising research directions.

\section{Fluid flow and deformation an the pore-scale}\label{sec:porescale_flow_deform}

\begin{figure} [h]
\centering
\includegraphics[width=0.95\columnwidth]{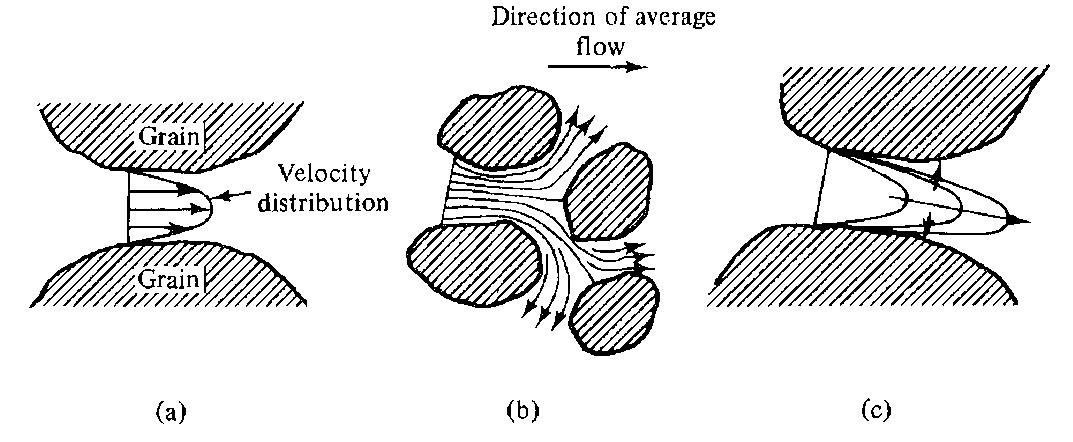}
\caption{Conventional depiction of hydrodynamic dispersion mechanisms operating at the pore scale (the graphic is reproduced from \cite{BearVerruijt:1987aa}): (a) mechanical dispersion via Poisueille flow profiles at pore throats, (b) mechanical dispersion via flow separation at obstacles, (c) spreading due to molecular diffusion (Brownian motion).}\label{fig:2Dstreamlines}
\end{figure}

In general, as fluid mixing and dispersion involves the interplay of fluid deformation, advection and molecular diffusion, successful upscaling of these phenomena must capture the salient features of these processes at the micro-scale. The classical picture of flow and transport at the pore scale (e.g. Bear and Verruijt, 1987 \cite{BearVerruijt:1987aa}) typically involves the depiction of fluid streamline dynamics meandering through a 2D domain punctuated by an arbitrary array of inclusions, as depicted in Figure~\ref{fig:2Dstreamlines}. This conceptual image illustrates the features which organise transport and dispersion but, if interpreted literally, the 2D nature of the fluid domain actually imposes severe constraints upon the admissible fluid dynamics, which in turn impact transport and mixing. Whilst conceptual, these topological constraints implicitly skew our understanding of the mechanisms which drive transport and mixing and neglect other important phenomena which may occur in 3D domains.

\begin{figure} [h]
\centering
\includegraphics[width=0.95\columnwidth]{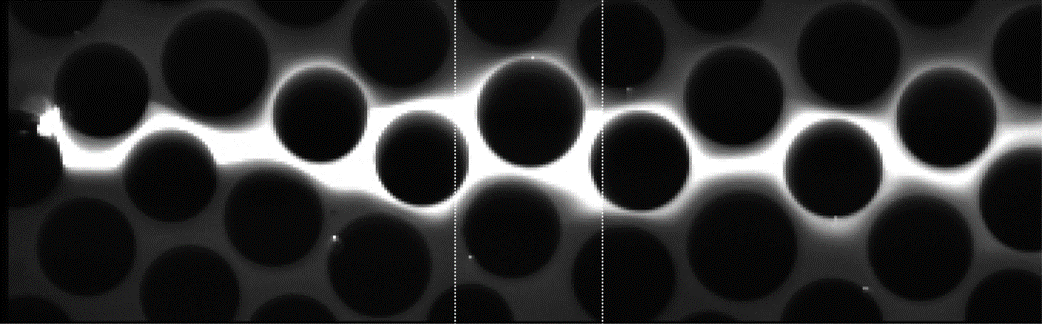}
\caption{Experimental image of hydrodynamic dispersion in 2D random porous media (mean flow is from left to right) illustrating splitting and recombination of fluid streamlines. Adapted from Bruyne et al~\cite{Bruyne:2014aa}.}\label{fig:2Dexperimental}
\end{figure}

For all steady 2D flows, an important topological constraint arises
from the fact that the fluid acts as a continuum, and so streamlines
cannot cross over each other, as reflected in
Figure~\ref{fig:2Dstreamlines}. As such, whilst the distance between
neighbouring streamlines in 2D flows may fluctuate along a streamline,
the separation distance cannot grow or shrink without bound due to
conservation of mass.  Although distance fluctuations can persist as
streamlines pass either side of the stagnation points of grains and
inclusions, eventually these streamlines must recombine, as shown in Figures~\ref{fig:2Dstreamlines},\ref{fig:2Dexperimental}.  This topological constraint, which is formally
imposed by the Poincar\'{e}-Bendixson
theorem~\cite{Teschl:2012aa,Balasuriya:2015}, places significant
restrictions on fluid deformation and hence transport of a diffusive
scalar (solutes, heat etc) since no steady 2D flow can be chaotic -
all such flows are regular with deformations which scale algebraically
with time. Hence the only mechanism for persistent dispersion of
solutes transverse to the mean flow direction is molecular
diffusion. That is, transverse diffusion is the only physical means by
which scalar quantities can spread transversely without bound in two
dimensions to generate the observed lateral spreading of macroscale
plumes. In contrast to the conventional picture of dispersion, the
divergence of streamlines around inclusions (specifically, around
stagnation points) is immaterial with respect to macroscopic
transverse dispersion as eventually these streamlines must recombine
downstream due to the topological constraint inherent to 2D steady
flow.

However, this topological constraint does not apply to 3D steady flows, as the additional degree of freedom associated with the third spatial dimension removes the restriction that a fluid streamline is bounded by its neighbours, and so streamlines may wander much more freely throughout the pore space. As such, continua may deform in a more complex manner, and streamlines may now diverge or converge without bound even if the flow is incompressible. Hence, 3D transport is not simply a 3D extrusion of the dynamics shown in Figure~\ref{fig:2Dstreamlines}, but rather 3D flows exhibit fundamentally different dynamics which introduce wholly new transport and mixing phenomena. It is these pore-scale dynamics that we shall explore and quantify in this study, and determine some of the related implications for macroscopic transport and dispersion.


\begin{figure}[tp]
\begin{centering}
\begin{tabular}{c}
\includegraphics[width=0.7\columnwidth]{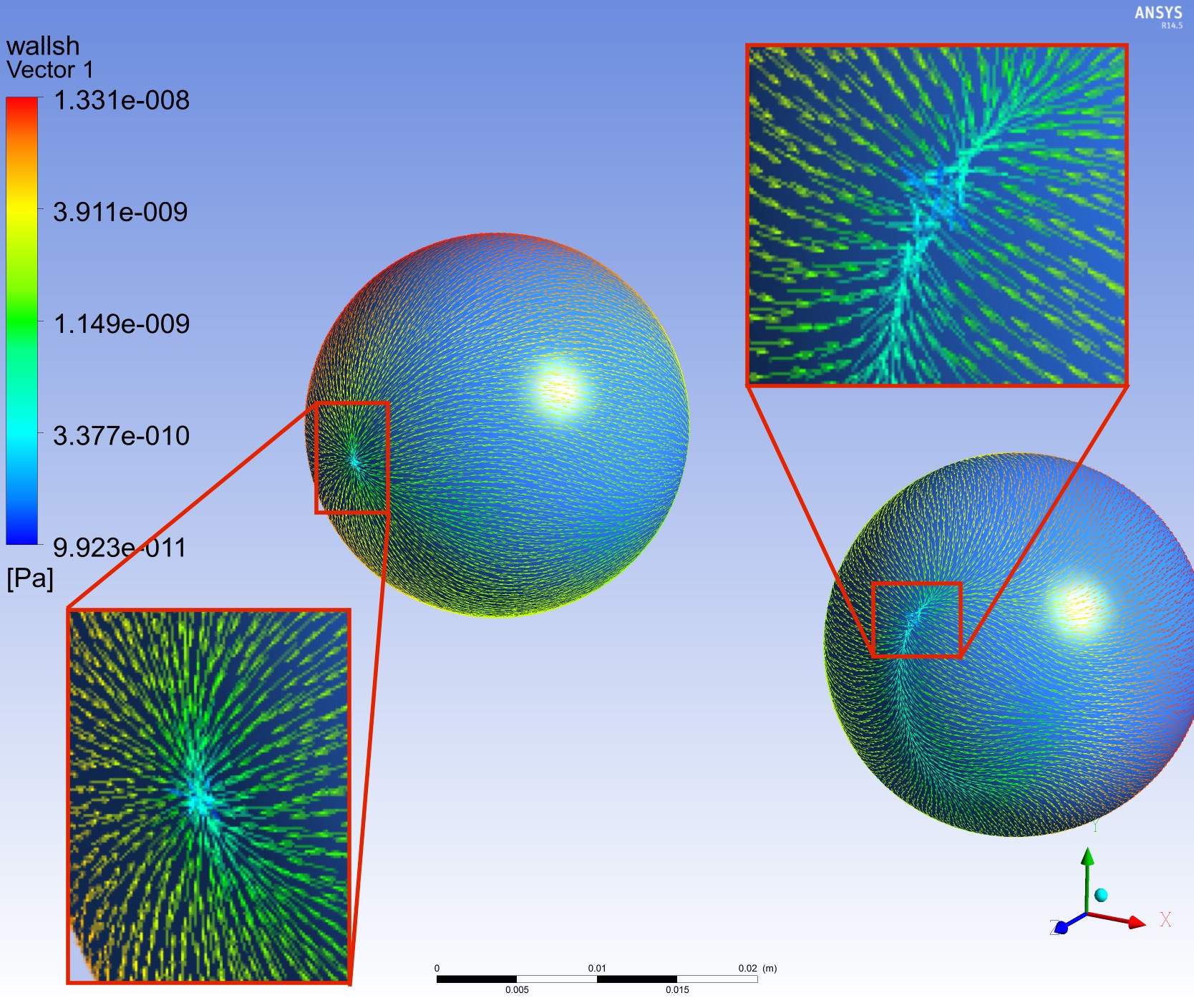} \\
(a) \\
\includegraphics[width=0.7\columnwidth]{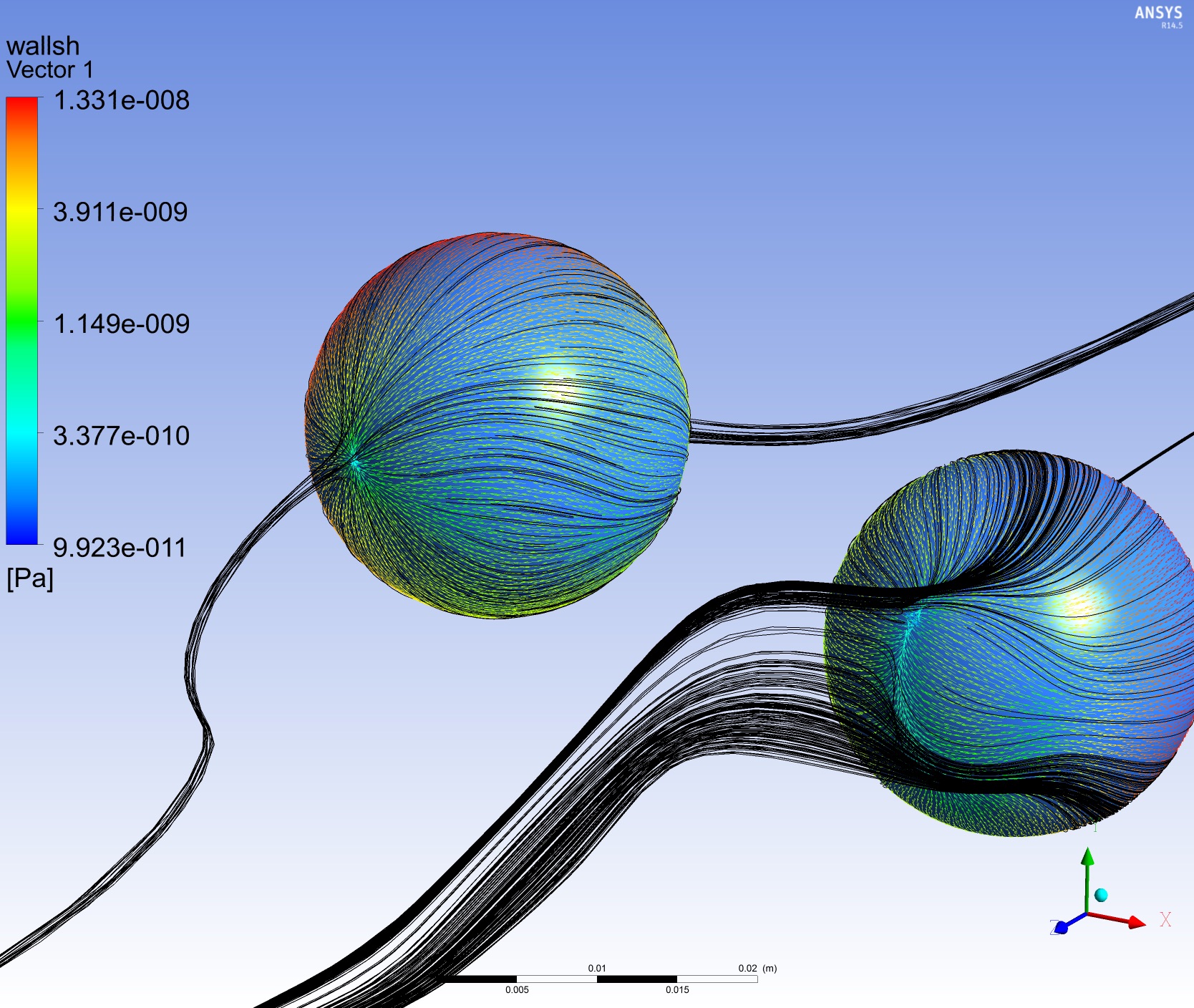}\\
(b)
\end{tabular}
\end{centering}
\caption{Computational simulations of 3D Stokes flow through a
  disordered array of spherical obstacles (only two spheres are
  shown). (a) Skin friction fields on sphere surfaces, showing minimum
  friction node point (left sphere) and locus (right sphere). (b)
  Panel (a) with streamlines drawn, showing a 1D unstable manifold
  downstream of the left sphere, and a 2D unstable manifold downstream
  of the right sphere. Image courtesy of Regis Turuban, University of Rennes.
  }\label{fig:3Dmanifolds}
\end{figure}

\begin{figure}[tp]
\begin{centering}
\begin{tabular}{c c}
\includegraphics[width=0.4\columnwidth]{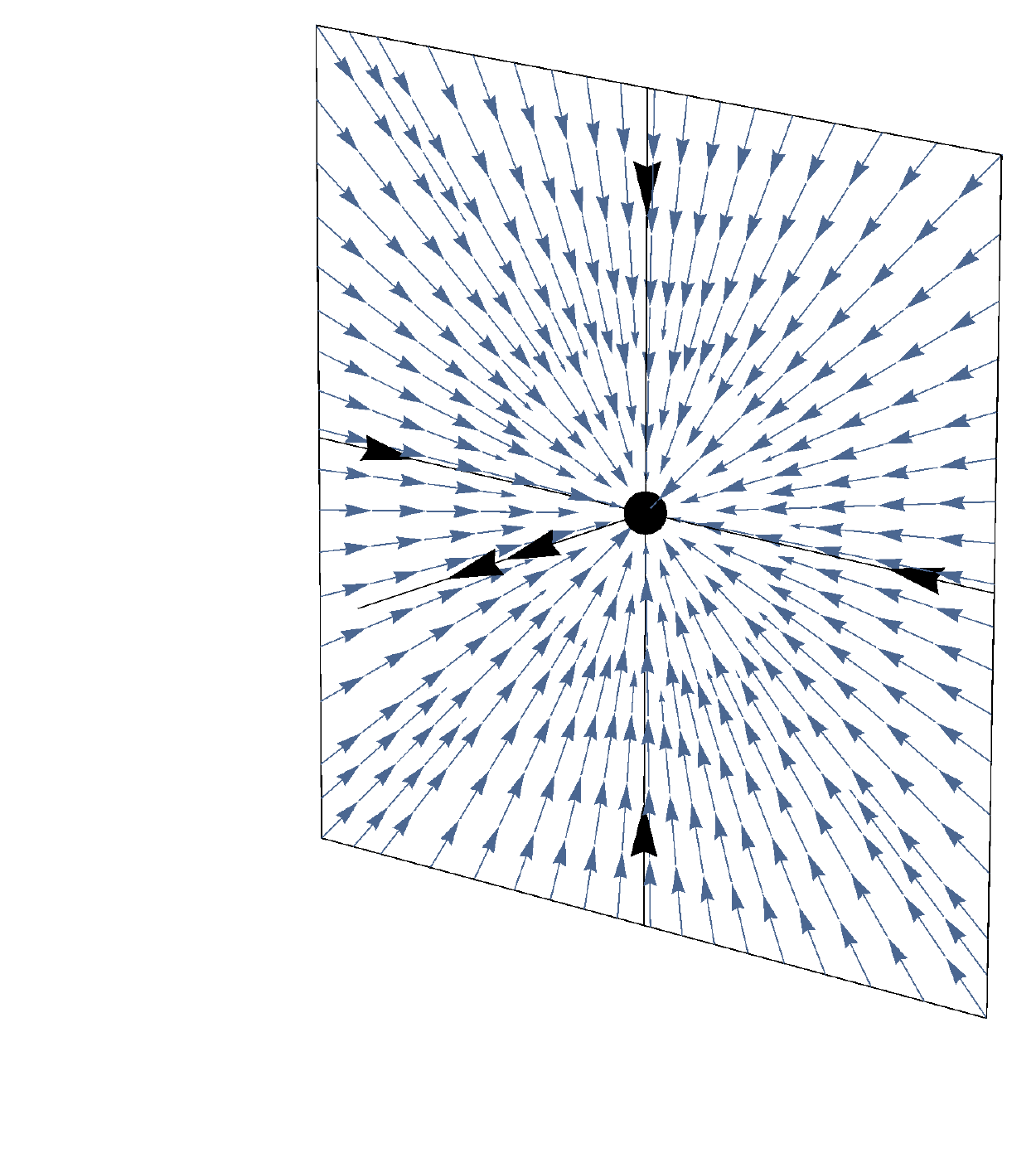} &
\includegraphics[width=0.4\columnwidth]{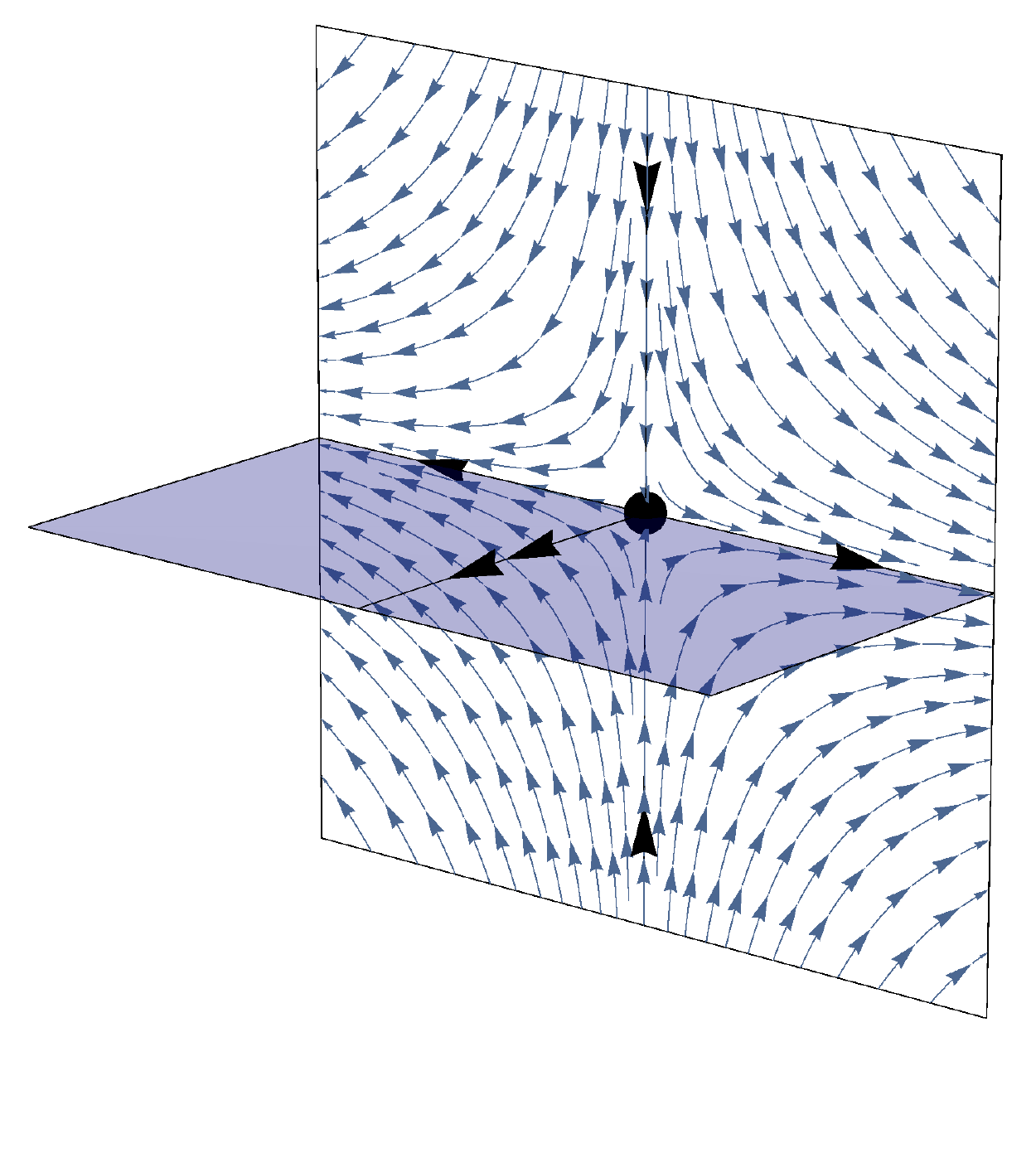}\\
(a) & (b)
\end{tabular}
\end{centering}
\caption{Schematic of the skin friction field (blue arrows) and 1D and 2D manifolds (black arrows, blue sheet) associated with an (a) node- and (b) saddle-type re-attachment point (black dot) shown in Figure~\ref{fig:3Dmanifolds}. }\label{fig:point_schematics}
\end{figure}

To understand the mechanisms which govern dispersion and dilution at
the pore scale, it is instructive to first consider the advection
dynamics in the absence of molecular diffusion; these dynamics
provides a template upon which diffusion, mixing and reaction play
out. Whilst pore-scale fluid flow may appear to be a trivial process,
in fact the Lagrangian advection dynamics can exhibit complex, chaotic
behaviour, even in simple steady flows such as Stokes
flow. Figure~\ref{fig:3Dmanifolds} shows the dynamics of a steady 3D
Stokes flow over a random cluster of non-touching spheres (many
spheres are present but only two are rendered),
with particular attention paid to streamlines which pass very closely
to the surface of the spheres shown and to the skin friction (surface
stress) field over the surface of these spheres. The complexity of
these streamlines is a direct consequence of the geometry of the
spheres; these inclusions significantly perturb the flow in a manner
which differs significantly from flow over an isolated
sphere. Figure~\ref{fig:3Dmanifolds} illustrates that the downstream
separatrix of the flow over a sphere can take the form of a point
(analogous to 2D flow in Figure~\ref{fig:2Dstreamlines}) or of a locus
which generates a 2D sheet (or manifold) of streamlines, even if the
upstream collection of streamlines is essentially one-dimensional
(1D).  Such stretching into 2D sheets is the fundamental building
block of chaotic advection at the pore scale, and we shall see that in
all disordered 3D geometries the repeated stretching into sheets is
the norm rather than the exception: such stretching is ubiquitous to
all random porous media.

The skin friction field itself causes transverse stretching of
streamlines as they are ejected from near the surface of the sphere,
and successive stretching events driven by these boundary dynamics
lead to complex entanglements of streamlines as fluid elements
encounter successive obstacles on the journey through a porous
medium. In contrast to steady 2D flows, streamlines in 3D flows that
are initially close are now free to diverge in such a manner that
fluid elements between streamlines can become stretched exponentially
with longitudinal distance without bound. As this unbounded
exponential growth is contained within a finite domain, these 2D
sheets are both stretched and folded into complex, highly striated
(lamellar) distributions as they flow downstream. Hence, even a
miniscule amount of molecular diffusion can lead to rapid mixing of
these lamellar structures and exhibit significantly accelerated
dilution, many orders of magnitude faster than that permitted by the
2D flows shown in Figure~\ref{fig:2Dstreamlines}. In this way
effective hydrodynamic dispersion can be far more prevalent and
profound in genuinely 3D steady flows.

The key questions with respect to dispersion and dilution are, given the topological freedom associated with 3D flow, how prevalent is such streamline separation/diversion and what are the consequent impacts upon transport and mixing? How do these mechanisms manifest at the macroscale and what are the implications of pore-scale dimensionality for macroscale dispersion and dilution?


\section{Chaos at the pore scale}\label{sec:pore_chaos}

As discussed in the previous Section, pore scale geometries and fluid
processes are both complex and complicated. We
contend, however, that by focusing on the essential topological
features of these systems it is possible to understand and classify
the dispersion and mixing dynamics without requiring precise
descriptions of the pore network. We pursue a detailed understanding
of the phenomena presented in Figure~\ref{fig:3Dmanifolds} and what
dynamical consequences arise for such flows encountering a long
succession of pore branches and merges. Whilst the phenomena in the
figure are the result of highly idealized models, we shall demonstrate
that the ingredients for such phenomena are inherent to all 3D random
porous media, regardless of the specific details of the pore-scale
architecture and so these mechanisms are completely ubiquitous.

To begin, the phenomena illustrated in Figure~\ref{fig:3Dmanifolds} are classical precursors of so-called \emph{chaotic advection} in a steady 3D flow. Chaotic advection~\cite{Aref:1984aa, Ottino:1989aa} refers to a particular flow phenomenon which can arise in unsteady 2D or steady 3D flows whereby the advection equation
\begin{equation}
\frac{d\mathbf{x}}{dt}=\mathbf{v}(\mathbf{x},t),\label{eqn:advection}
\end{equation}
describing the evolution of the position $\mathbf{x}$ of a fluid
particle with time $t$ under the action of the flow field
$\mathbf{v}(\mathbf{x},t)$ can give rise to chaotic dynamics in which
fluid particles separate exponentially in time, losing memory of their
initial neighborhoods and giving rise to complex fluid particle
trajectories.  The strength of the chaotic dynamics is characterized
by the Lyapunov exponent $\lambda_\infty$ which measures the
exponential growth rate of the length $\ell(t)$ of a fluid element of
initial length $\ell_0$ with time as
\begin{equation}
\lambda_\infty=\lim_{t\rightarrow\infty}\frac{1}{t}\ln \frac{\ell(t)}{\ell_0}.\label{eqn:Lyapunov}
\end{equation}
The Lyapunov exponent also acts as an indicator for chaotic dynamics
as sub-exponential (i.e. algebraic) stretching results in
$\lambda_\infty=0$. Due to conservation of mass, exponential
longitudinal stretching of fluid elements must also be accompanied by
equivalent exponential narrowing in at least one of the transverse
directions, resulting in highly striated lamellar material
distributions. To contain such elongated structures within a finite
domain, both fluid stretching and folding is required, and these two
motions are considered to be the hallmarks of chaotic dynamics in all
continuous systems~\cite{Ottino:1989aa}.

This stretching and folding process is known as chaotic advection, and
represents an efficient means of fluid stirring akin to that of
turbulent flow but available for acctuation in low Reynolds number,
non-turbulent flows.  The key difference between chaotic advection and
turbulent mixing is that the advection equation (\ref{eqn:advection})
is an identity and so is kinematic rather than dynamic in origin, and
so chaotic advection can arise under a broad range of dynamical
regimes. Importantly, from (\ref{eqn:advection}), flows which have a
simple structure in the Eulerian frame can generate complex Lagrangian
dynamics and material distributions with arbitrarily small length
scales.  Since at sufficiently small length scales molecular
diffusion---no matter how weak---plays a dominant role and leads to
rapid dilution and homogenization of initially large-scale
concentration distributions, chaos is the basis for accelerated
mixing.  This phenomenon is well
documented~\cite{Ottino:1989aa,Wiggins:2004aa,Sundararajan:2012aa,Metcalfe:2012aa}
and understood in low-Reynolds number flows such as Stokes and laminar
flows, and forms the design basis of a wide range of
micro-fluidic~\cite{Stroock:2002aa} and
industrial~\cite{Lester:2009aa} mixing devices, including static
mixers \cite{Hobbs:1997aa} which operate on similar principles to the
mixing dynamics in 3D porous media. Moreover, elucidation of these
mixing mechanisms has provided significant insights into mixing and
transport in nature, ranging from transport in geophysical
flows~\cite{Mezic:2010aa} to population biology~\cite{Karolyi:2000aa}.

It is instructive to consider whether the mechanism of chaotic advection is relevant to transport in porous media; particularly with respect to dilution, mixing and macro-dispersion. For reasons that shall become apparent, we focus attention to pore-scale flow, under steady 3D Stokes flow, subject to no-slip boundary conditions. Whilst the concepts discussed apply to broader conditions than these (specifically any boundary condition and all steady flows of continua), these conditions serve well to elucidate the fundamental mechanisms.

\begin{figure} [h]
\centering
\includegraphics[width=0.8\columnwidth]{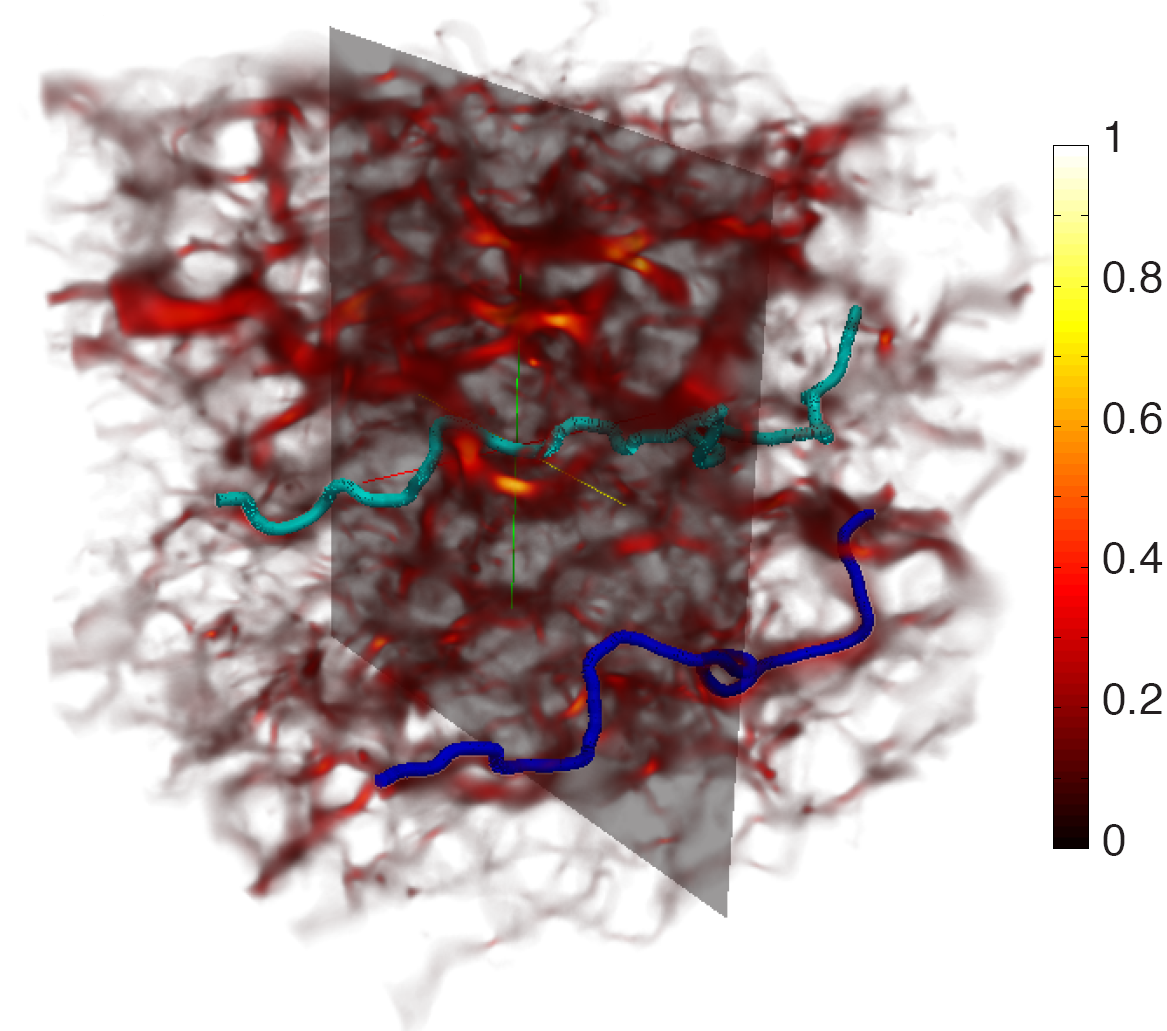}
\caption{Digital reconstruction of the 3D flow field through a Berea
  sandstone block (1.6 mm cube).  The normalized velocity magnitude is
  rendered according to the colour bar; yellow zones indicate high
  velocity bursts.  Two flow paths (blue and cyan) show evidence of
  branch and merge behaviour.  Adapted from Figure~1 of
  \cite{Kang:2015}.}\label{fig:3Dpoland}
\end{figure}

The key feature common to all porous media is topological complexity
of the pore space, such that the pore-scale fluid domain is highly
connected, and contains a large number of junctions and merges per
unit volume, as illustrated by the blue and cyan flow paths in Figure~\ref{fig:3Dpoland}). This defining property is
common to all porous media, whether open porous networks, granular,
fractured or synthetic media and has significant implications for flow
and transport. Such complexity can be quantified via digital CT
imaging of the pore space, allowing direct measurement of the number
of pores $N$, redundant connections $C$ and enclosed cavities $H$ per
unit volume~\cite{Vogel:2002}. These measures allow the connectivity
at the pore scale to be directly quantified in terms of the
\emph{topological genus} $g$, which may be interpreted as the number
of distinct ``holes'' an object possesses, such that a sphere is of
genus zero, whilst a doughnut is of genus one, and a pretzel may be
genus two, three or higher. The topological genus per unit volume of a
porous medium is then directly related to the measures
above~\cite{Vogel:2002} as
\begin{equation}
2(1-g) =N-C+H.\label{eqn:Eulerchar}
\end{equation}
As all porous media are highly connected (topologically complex) at
the pore scale, imaging studies typically find $N < C$ and $H$ small (except for foams and some vitreous
rocks), and so $g$ is positive and large across a broad range of systems.  This inherent topological
complexity has significant implications for flow and transport at the
pore-scale.

Remarkably, there exists a direct correspondence between pore-scale
complexity (measured by $g$) and the number and type of stagnation
points in the skin friction vector field (the shear stress field
projected onto the pore boundary), such as those shown in
Figure~\ref{fig:3Dmanifolds}. A stagnation point $\mathbf{x}_p$ arises
as a zero of the skin friction field $\boldsymbol\tau(\mathbf{x})$,
and may take the form of a \emph{separation point} associated with the
separation of fluid from the boundary region into the fluid bulk, or a
\emph{reattachment point} associated with the converse attachment of
fluid onto the boundary region.  Attachment points predominantly lie
on upstream facing surfaces, whilst separation points typically lie on
downstream facing surfaces.

Both separation and reattachment stagnation points may be further
classified as either node or saddle points, in reference to the
dynamics of the skin friction field local to $\mathbf{x}_p$. These
types are quantified by the index $\gamma_p$, where $\gamma_p$=-1 for
a saddle point, $\gamma_p$=+1 for a node point. Typical node and
saddle points and the associated skin friction fields from computational studies are shown in Figure~\ref{fig:3Dmanifolds}, and schematics of these features are illustrated in Figure~\ref{fig:point_schematics}. The sum of these indices is directly related to the
topological genus $g$ by the Poincar\'{e}-Hopf theorem
\begin{equation}
\sum_p \gamma_p(\mathbf{x}_p)=2(1-g),\label{eqn:PHopf}
\end{equation}
forming the direct link between the topology of the pore space and the
dynamics of the skin friction field.  It is important to note that
this link is not related to the pore-scale fluid mechanics per se, but
rather is a direct consequence of the advection of a continuum through
the highly connected pore space.  As explained below, saddle points
are important for the generation of chaotic advection because they not
only impart significant fluid stretching, but also because the
stretching surfaces (termed unstable manifolds in the language of
dynamical systems) that emanate into the fluid bulk from saddle points
are 2D, whereas node points lead to 1D stretching lines in the flow
field.

The impact of these different points is illustrated in Figure~\ref{fig:3Dmanifolds}, which shows the behaviour of a family of streamlines which travel closely to the surface of a sphere prior to being ejected into the fluid bulk near a separation point. If the separation point is of node type, the family of streamlines are tightly bound around a 1D line (a 1D unstable manifold), whereas if the separation point is of saddle type, the streamlines align with a 2D surface (a 2D unstable manifold). As either a 1D line or a 2D surface local to a stagnation point is stretched exponentially by the flow, then local fluid trajectories are also attracted toward these structures due to conservation of volume. The behaviour is reflected by the shadowing behaviour of the streamlines in Figure~\ref{fig:3Dmanifolds}.

\begin{figure}[h]
\centering
\includegraphics[width=0.8\columnwidth]{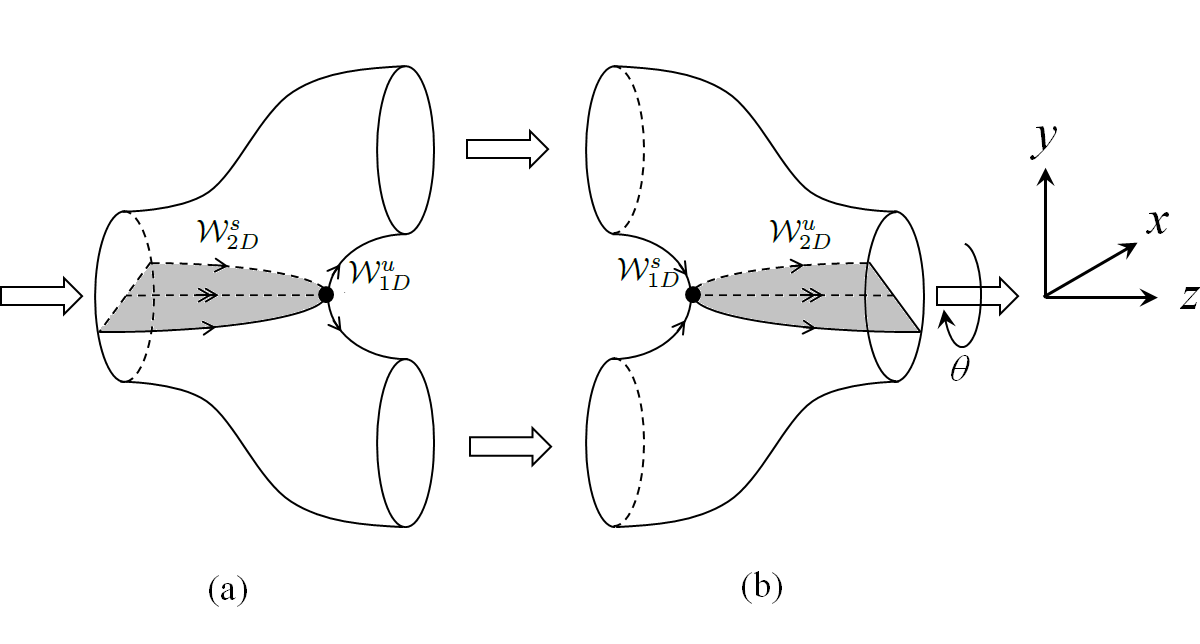}
\caption{Schematic of pore branch (a) and merge (b) depicting stagnation points and associated 2D unstable $\mathcal{W}^U_{2D}$ and stable $\mathcal{W}^S_{2D}$ manifolds which act as surfaces of transverse minimum transverse flux. Note fluid stretching transverse to these surfaces and the orientation angle $\theta$ between these surfaces. Adapted from~\cite{Lester:2013aa}}.\label{fig:pore_branch}
\end{figure}

This stretching also means that the 2D surface associated with a
saddle point is a \emph{surface of minimal transverse
  flux}~\cite{MacKay:1994}, in that transport across this surface is
significantly retarded, and so such surfaces form largely impenetrable
barriers which divide and separate the 3D flow field as shown in
Figure~\ref{fig:pore_branch}. Conversely, whilst the 1D lines
associated with node points impart fluid stretching, this effect is
local to a small region around the 1D line, and other fluid elements
can travel around this line relatively unperturbed by its presence. Similarly, saddle-type attachment points also admit 2D
surfaces of minimal transverse flux (as per Figure~\ref{fig:pore_branch}(a)), and node-type attachment points
admit 1D lines, however both of these are contracting rather than
stretching structures (termed 1D and 2D stable manifolds). Again,
these surfaces impart significant fluid stretching, but it is the
interaction of the 2D stretching and contracting surfaces which
governs the persistence of these deformations.

Chaotic advection in steady 3D flow arises when the 2D stable and unstable manifolds associated with saddle-type separation and reattachment points respectively intersect each other transversely (i.e. $\theta\neq 0$ in Figure~\ref{fig:pore_branch}), leading to persistent exponential fluid stretching. Conversely, if these surfaces intersect each other tangentially ($\theta=0$ in Figure~\ref{fig:pore_branch}), the stretching and contraction cancel each other out, leading to significantly reduced net fluid deformation. In steady 2D flows, $\theta$ is constrained to be zero always, prohibiting persistent exponential fluid stretching. 

\begin{figure}[h]
\centering
\includegraphics[width=0.85\columnwidth]{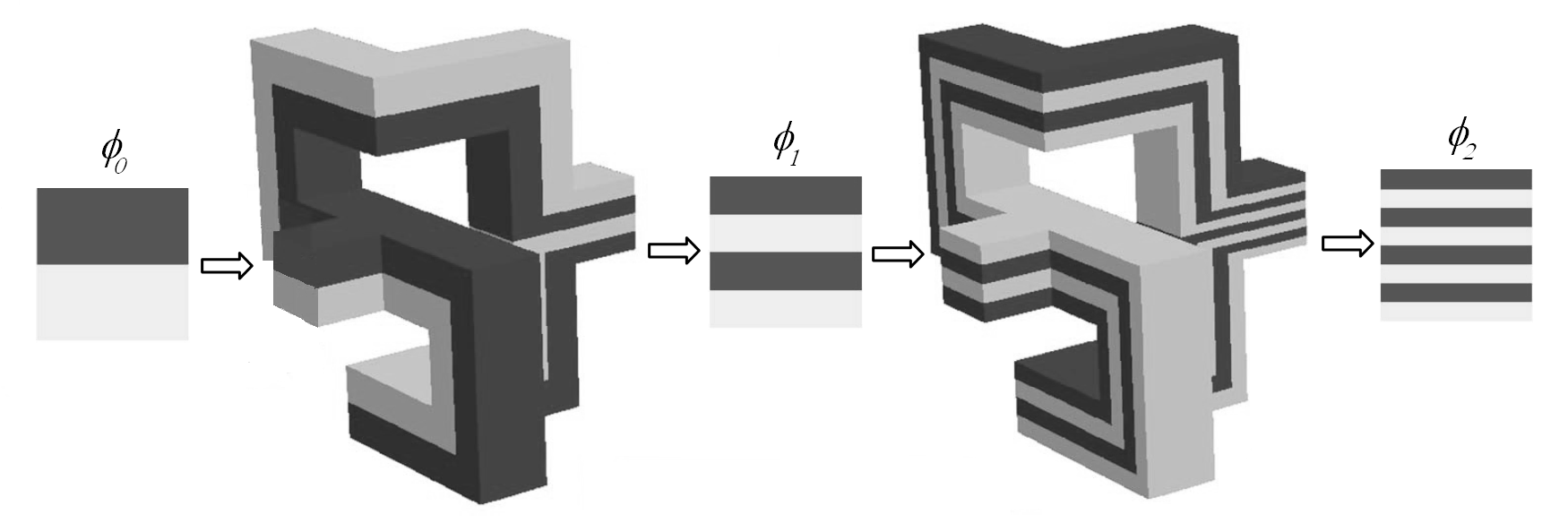}
\caption{Schematic of the baker's flow, a 3D fluid mechanical implementation of the baker's map arising from non-trivial pore branching and merging. Adapted from~\cite{Carriere:2007aa}.}\label{fig:3Dbakermap}
\end{figure}

These stretching dynamics are shown schematically in the 3D baker's
flow depicted in Figure~\ref{fig:3Dbakermap}, adapted from
\cite{Carriere:2007aa}, which corresponds to the case $\theta=\pi/2$.
This schematic depicts fluid deformation (as illustrated by the
dark/light bands) over a pore branch and subsequent merger, followed
by a repeated branch and merger, under steady flow conditions. Whilst
not shown, saddle-type stagnation points occur at each of the pore
branches and mergers, and the associated 2D unstable manifolds project
into the fluid bulk oriented at $\theta=\pi/2$ to each other, where
the ``twist'' between the branch and merger is necessary to achieve
this angle. The input flow (at left, $\phi_0$) is partitioned into two
(light and dark) tones; after passing through two successive branch
and merger re-arrangements, the fluid tones are re-distributed into an
array of eight contiguous light and dark tones (at right, $\phi_2$),
indicating that significant stretching and folding process has taken
place along the flow trajectory. As per Figure~\ref{fig:3Dbakermap},
the 3D baker's flow generates 2D fluid striations (sheets) which are
stretched and thinned exponentially with the number $n$ of pore
branches and merges as $2^n$.  Numerical
simulations~\cite{Carriere:2007aa} of Stokes flow in this geometry
calculated that the Lyapunov exponent $\lambda_\infty\approx 0.68$ for
this flow is withing 2\% of the theoretical upper bound
$\lambda_\infty=\ln 2$ for continuous steady 3D flows.  Hence fluid
stretching arising from pore branches and merges in certain geometries
can generate very efficient mixing from chaotic advection.  Note that
these Lyapunov exponents are calculated with respect to the number $n$
of coupled pore branches and mergers rather than the time $t$ in
(\ref{eqn:Lyapunov}).

\begin{figure}
\begin{centering}
\begin{tabular}{c c}
\includegraphics[width=0.45\columnwidth]{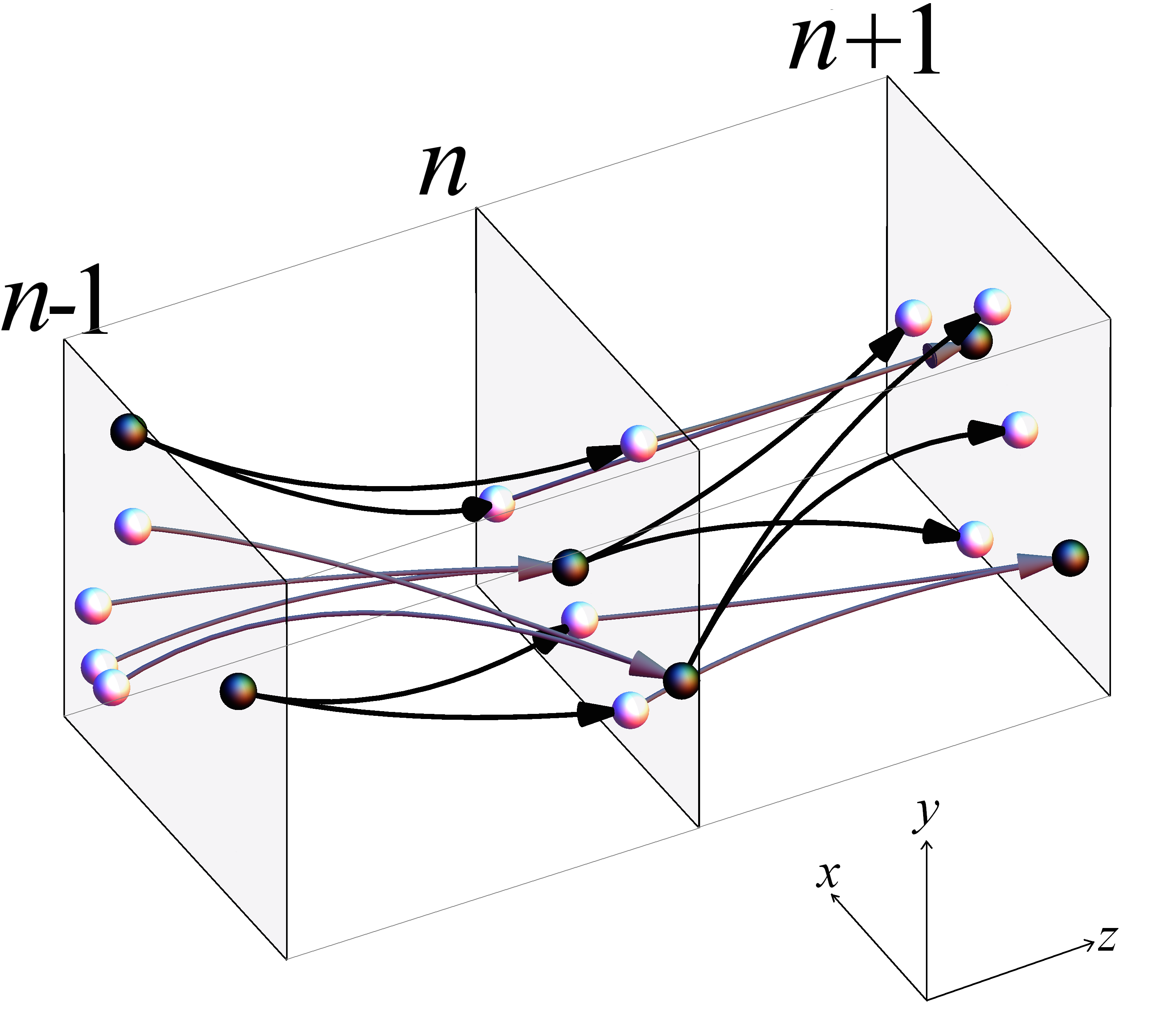}&
\includegraphics[width=0.45\columnwidth]{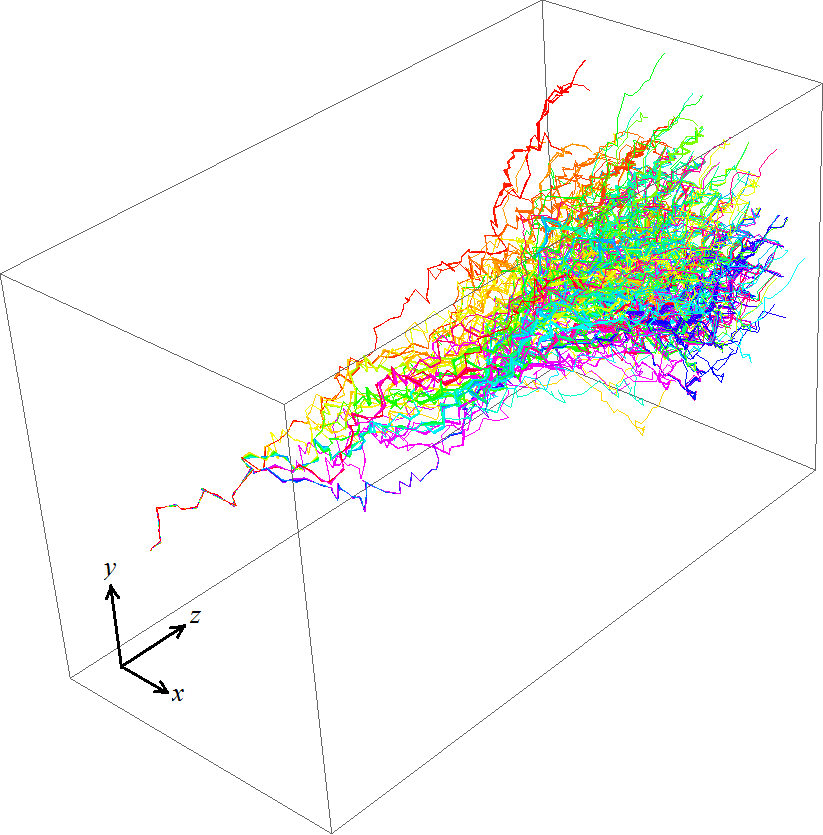}\\
(a) & (b)
\end{tabular}
\end{centering}
\caption{(a) Schematic of connectivity of pore branches (black) and
  mergers (white) between mapping planes in the open porous network
  model of \protect\cite{Lester:2013aa}, with branch pores (black) and
  merge pores (white) shown. (b) Evolution of a continuously injected
  line source of differently colored particles in a single pore within
  one realisation of the open porous network
  model.}\label{fig:pore_network}
\end{figure}

\begin{figure}
\begin{centering}
\begin{tabular}{c c c c}
\includegraphics[width=0.22\columnwidth]{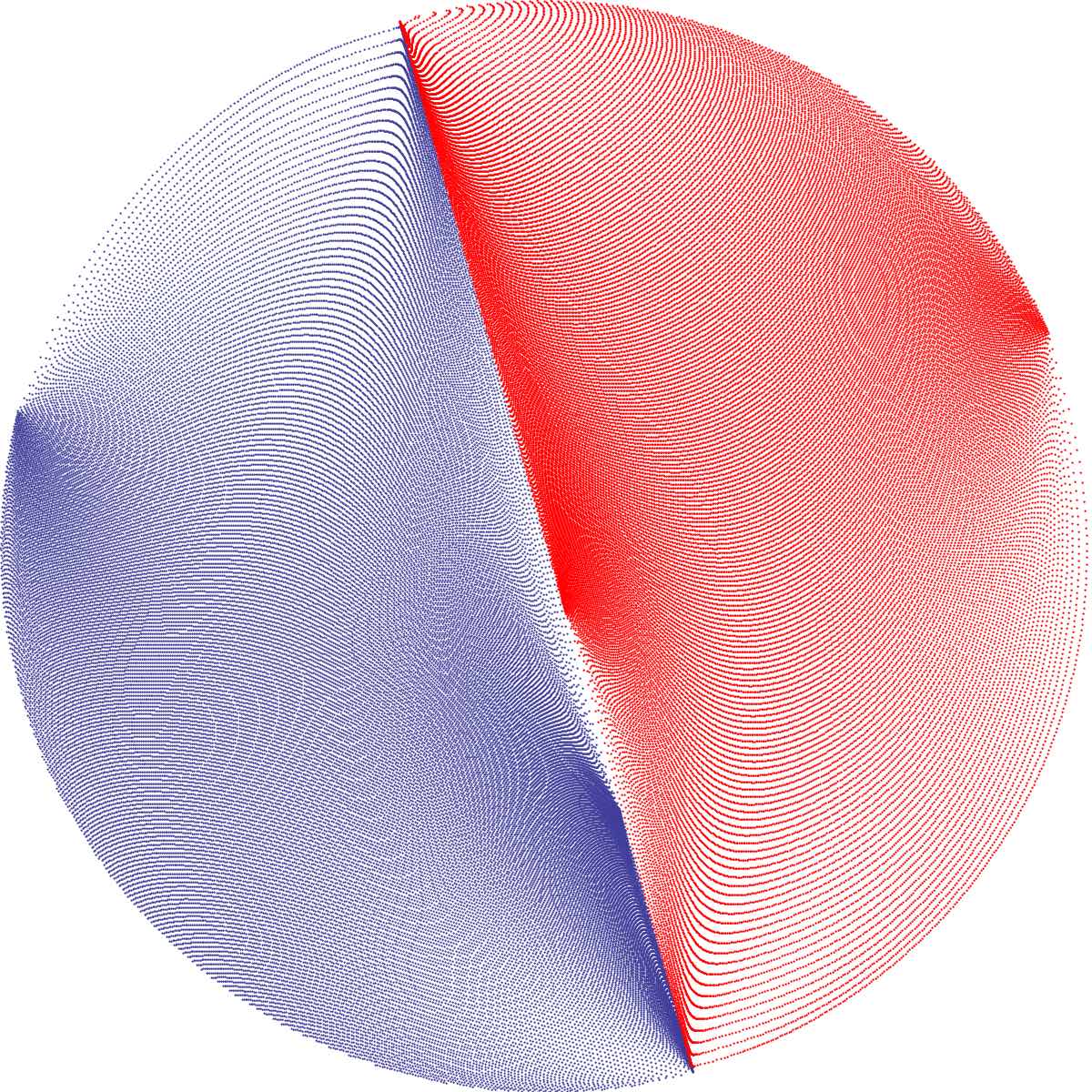}&
\includegraphics[width=0.22\columnwidth]{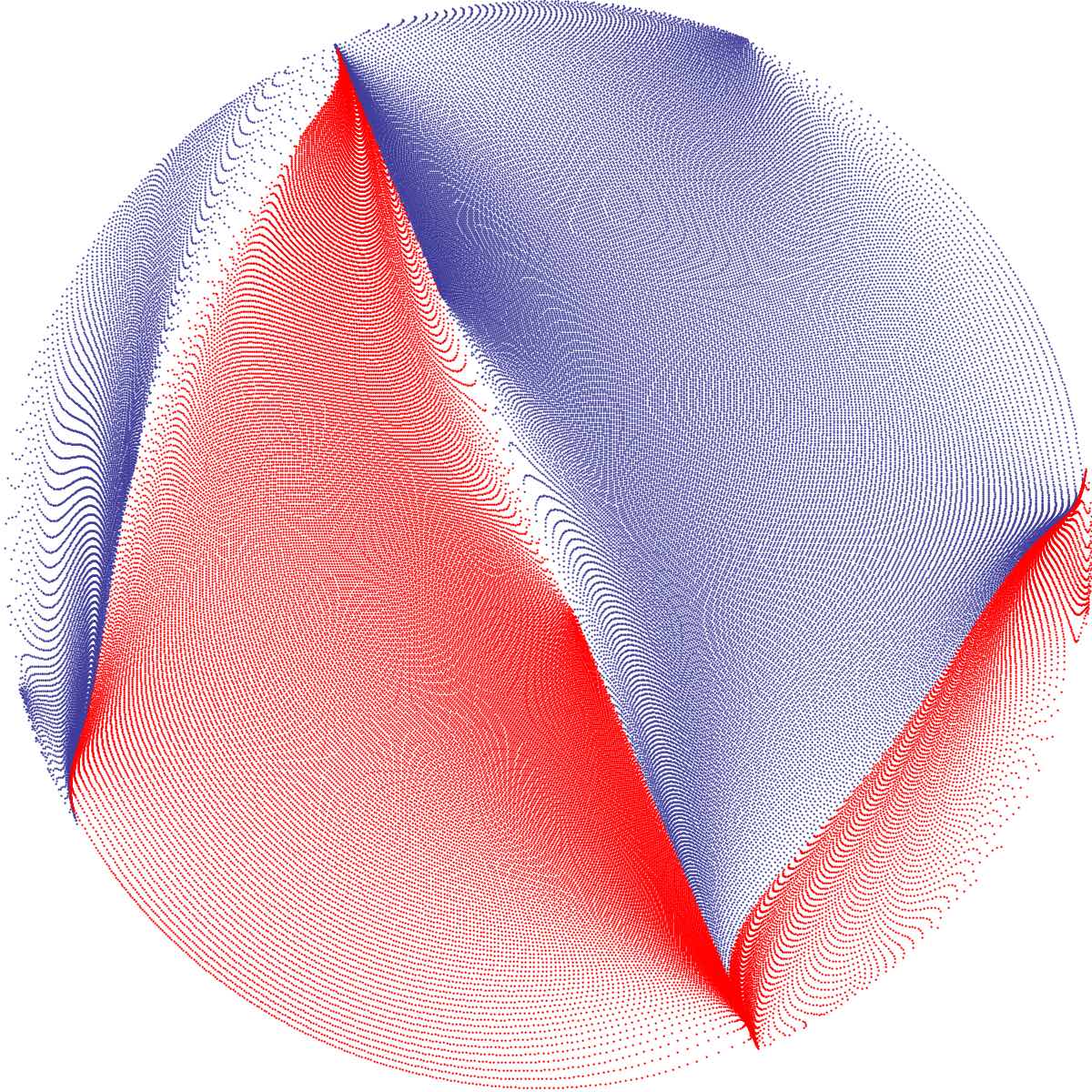}&
\includegraphics[width=0.22\columnwidth]{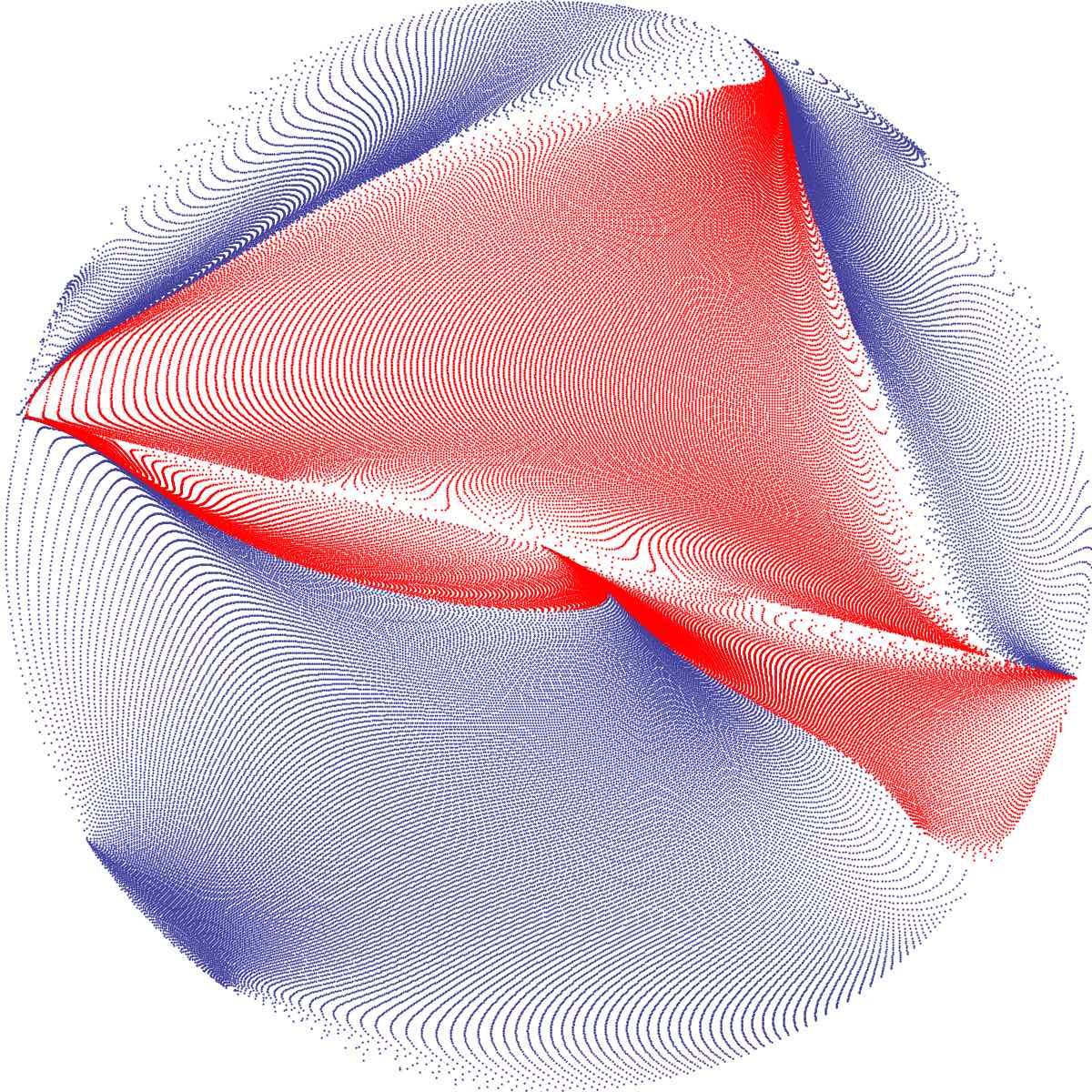}&
\includegraphics[width=0.22\columnwidth]{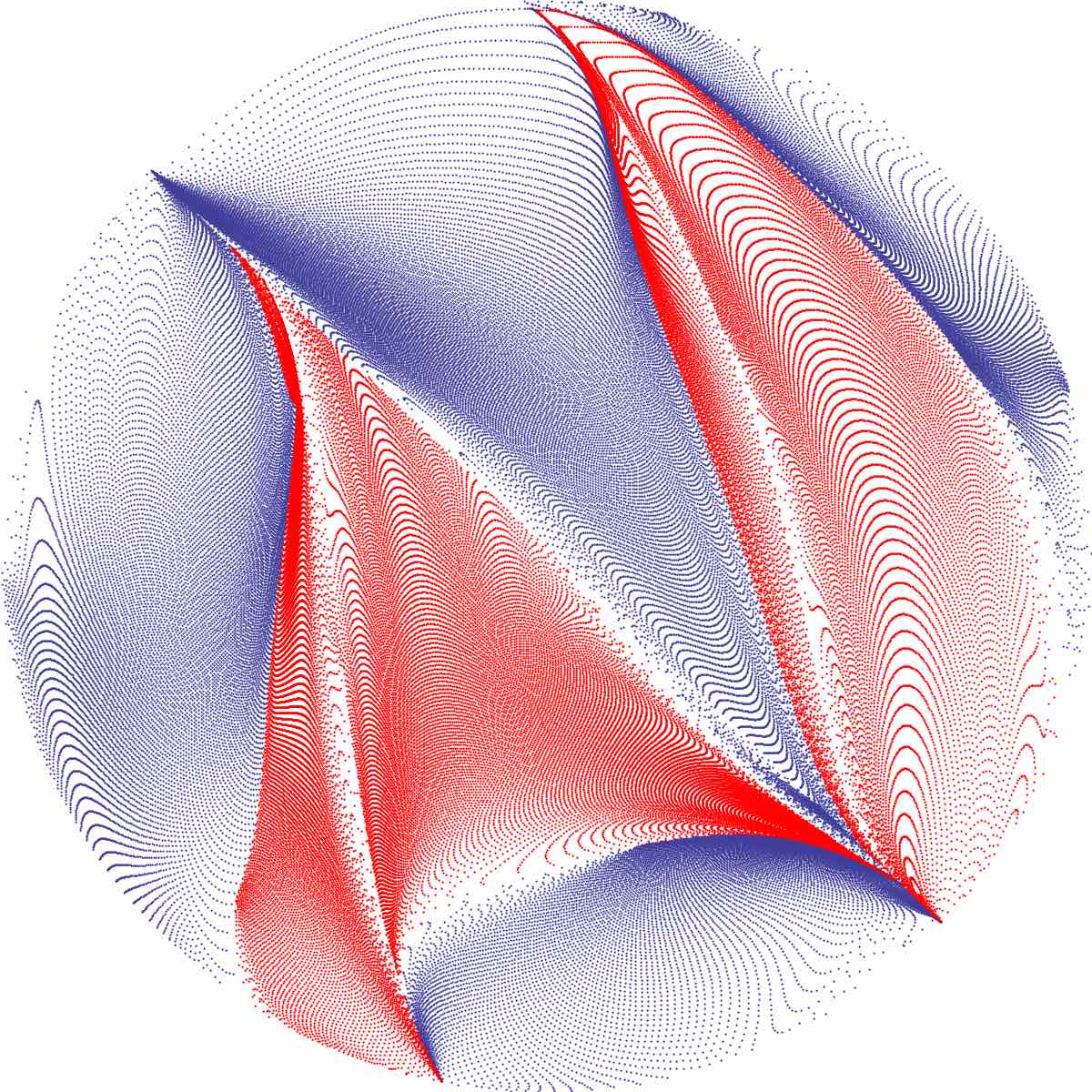}\\
$n=1$ & $n=2$ & $n=6$ & $n=8$\\
\includegraphics[width=0.22\columnwidth]{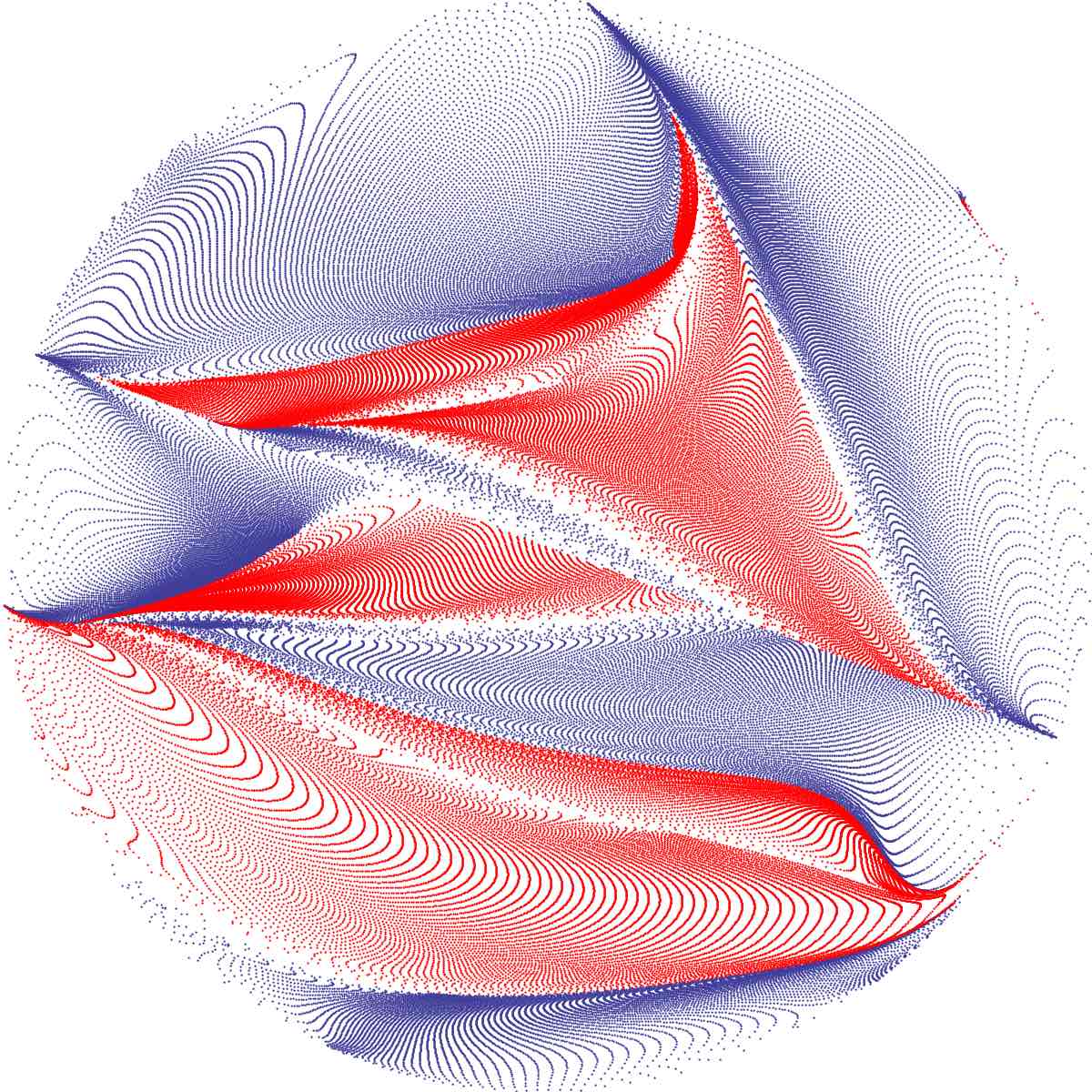}&
\includegraphics[width=0.22\columnwidth]{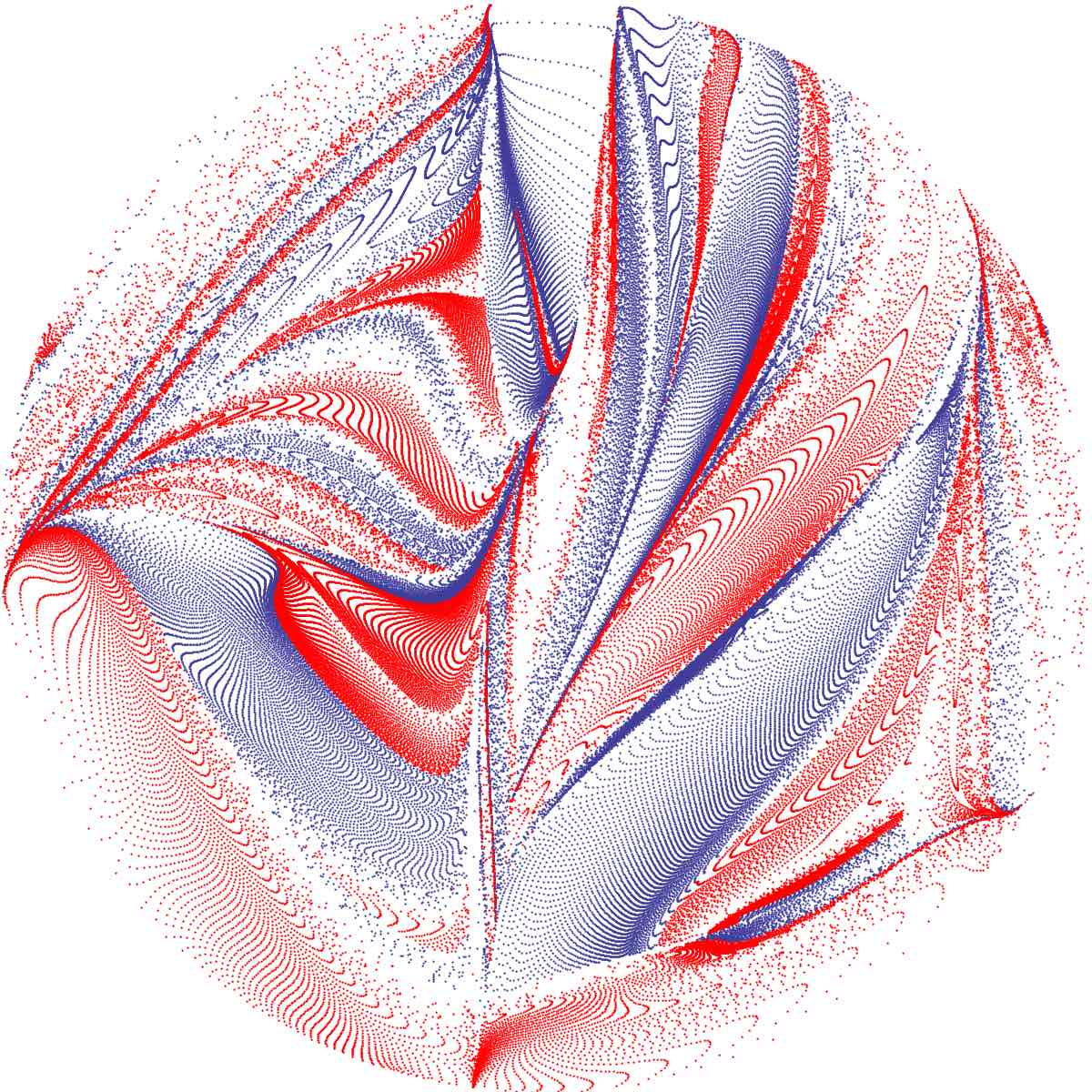}&
\includegraphics[width=0.22\columnwidth]{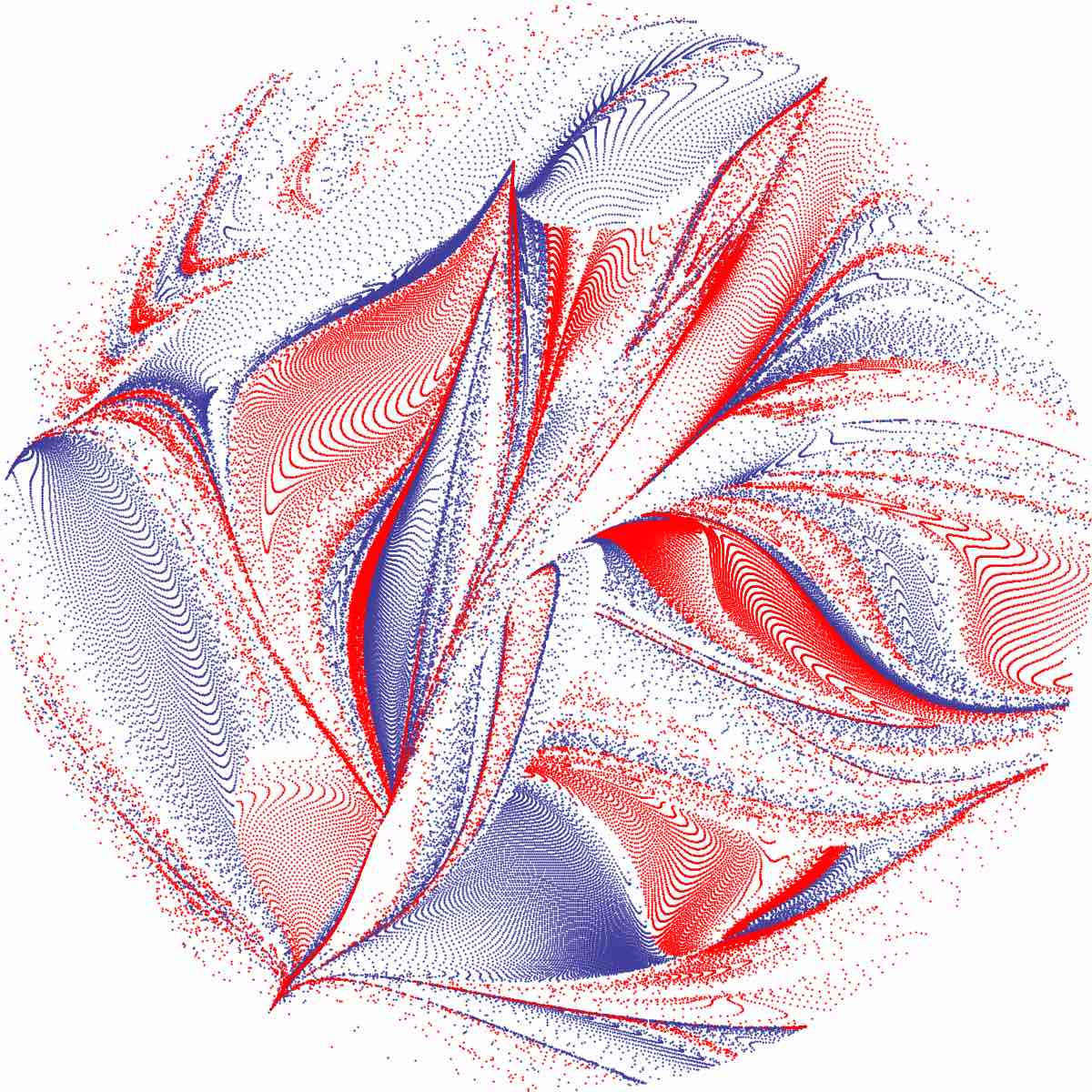}&
\includegraphics[width=0.22\columnwidth]{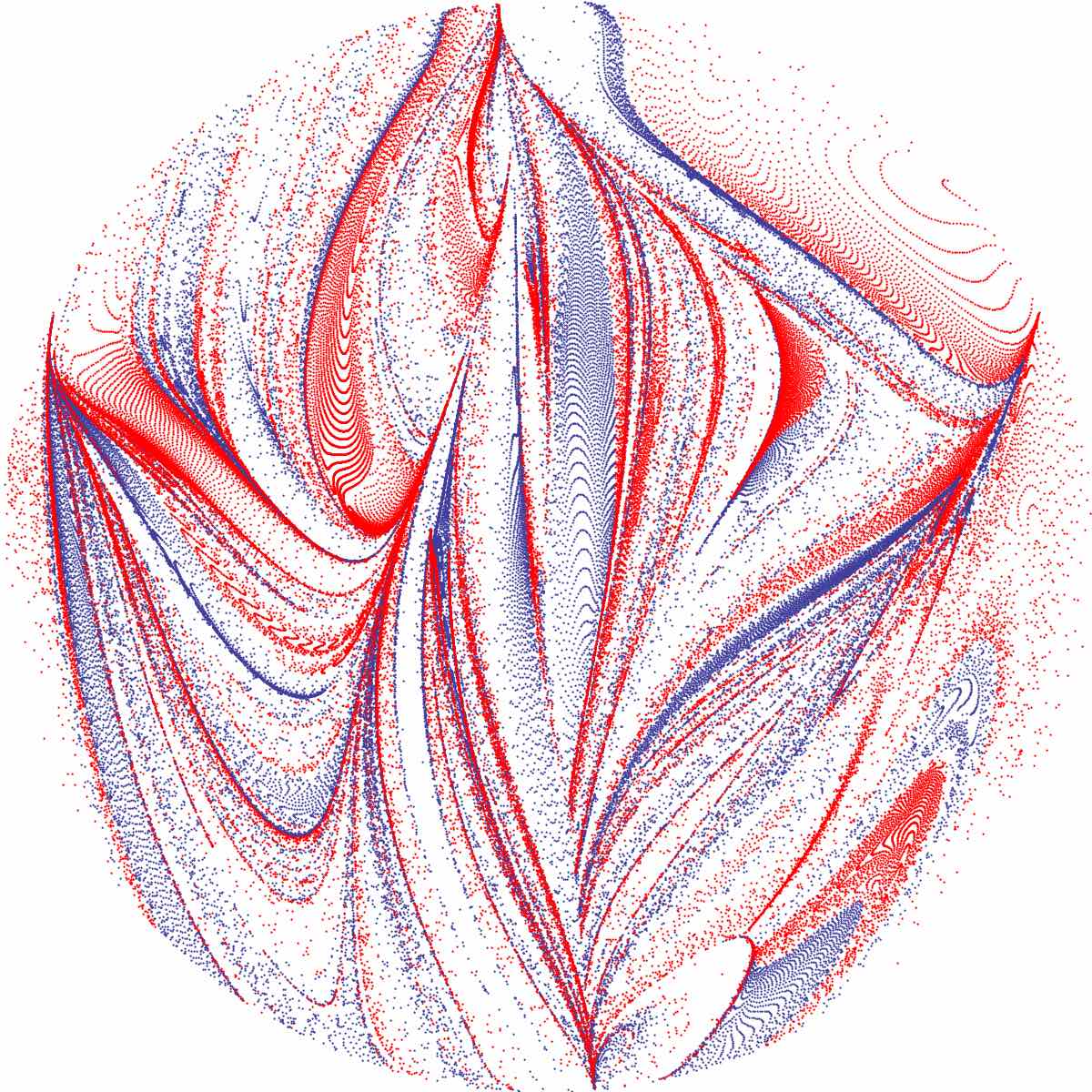}\\
$n=10$ & $n=20$ & $n=30$ & $n=40$
\end{tabular}
\end{centering}
\caption{Typical evolution of coloured fluid particles with pore number $n$ in a 3D randomly oriented open porous network.}\label{fig:dyetrace}
\end{figure}

For random porous media, the orientation angle $\theta$ is not at the
optimum value $\theta=\pi/2$, but rather is randomly distributed
(uniformly as an independent random variable) in the range $[0,\pi]$
between branches and mergers.  For a random open porous network
(detailed in Section~\ref{sec:network_model}) comprising randomly
connected pore branches and mergers, the mean stretching rate was
calculated in \cite{Lester:2013aa} to be
$\lambda_\infty\approx 0.1178$.  A schematic of the model connections
in \cite{Lester:2013aa} is shown in Figure~\ref{fig:pore_network}(a),
and a typical macroscopic evolution of a dye plume continuously
injected into a single pore is shown in
Figure~\ref{fig:pore_network}(b).  Chaotic advection inherent to this
open porous network rapidly organises these non-diffusive particles
into highly striated material distributions which grow exponentially
as $\ell(n)=\ell_0\exp(\lambda_\infty n)$ as shown in
Figure~\ref{fig:dyetrace}, which illustrates the cross-sectional
distribution of fluid tracer particles in a single pore as they evolve
with pore number $n$ under a mean advective flow. Whilst these
striations do not stretch as rapidly as for the 3D bakers flow shown
in Figure~\ref{fig:3Dbakermap}, the underlying mechanism leading to
chaotic advection in the random porous network is identical, and is
also intrinsically applicable to granular systems and fracture
networks.

The ergodicity of streamline dynamics under chaotic advection means that the Lyapunov exponent $\lambda_\infty$ reflects both an ensemble average as well as a volumetric average. Although it may be surprising that fluid stretching in random porous media can lead to persistent exponential fluid stretching (as random stretching and compression may be expected to cancel), this persistence is explained by the asymmetry between fluid stretching and compression. Whilst material elements tend to align with the stretching direction and so amplify stretching, under compression material elements tend to align normal to the compression direction, hence compression is retarded. The Lyapunov exponent $\lambda_\infty\approx 0.1178$ directly quantifies this asymmetry.

Whilst significantly smaller than the upper bound
$\lambda_\infty=\ln 2$, this mean stretching rate is still positive,
which means that the topological complexity inherent to all porous
media imparts significant chaotic advection to all 3D random media
under steady flow conditions.  As each branch and merge
  pair corresponds roughly to one pore (exactly in the model) and as
  significant mixing and spreading happens after a few tens of pore
  lengths (i.e. $l/l_0 > 100$ for $\lambda_\infty \approx 0.1178$ for
  $n > 39$ pores), mixing and dispersion by chaotic advection is significant
  well within the averaging volumes typically associated with
  upscaling models: dispersion by chaotic advection cannot rightly be
  ignored when upscaling. Although the model used to develop these quantitative estimates is
highly simplified, the fundamental properties (topological complexity
and random pore orientation) are inherent to all 3D random porous
media. Moreover, this rate presents a lower bound driven by the
pore-space topology, and so does not account for for additional
mechanisms such as surface roughness~\cite{Stroock:2002aa} and pore
curvature~\cite{Jones:1989aa}. Although fluid shear is also
significant in all porous media (due to the important role played by
the no-slip condition at grain surfaces), such mechanisms give rise
only to algebraic fluid deformations, and so are dominated in the long
term by the exponential fluid stretching associated with chaotic
advection.

In this Section we have seen that there exist significant qualitative differences between the fluid dynamics in 2D and 3D porous media. In the latter case, complex streamline behaviour driven by the pore-scale topology generates chaotic advection, resulting in highly striated, lamellar material distributions which have the potential to accelerate pore-scale mixing and augment macroscopic transport. In the following Sections we explore the impact of these dynamics upon pore-scale transport and mixing which manifest as macroscopic dispersion and dilution.

\section{Flow and transport in a model open porous network}\label{sec:network_model}

To study the fundamentals of chaotic advection and transport in 3D
random porous networks, we consider an idealised random model network
based on the symmetric pore branch element $\Omega$ (shown in
Figure~\ref{fig:pore_branch}(a)) which contains the minimum
topological complexity and disorder common to all random porous media,
\textit{i.e.} branching and merging pores. This ideal network is
homogeneous and isotropic at the macroscale, loosely defined as the
length scale significantly greater than the characteristic length
scale of $\Omega$. Whilst real porous media exhibit significantly more
complexity in the pore-scale structure including features such as
tortuosity, heterogeneity and surface roughness, for clarity of
exposition we seek the simplest model which exhibits chaotic advection
and anomalous transport. These phenomena persist in the presence of
additional pore-scale features outlined above~\cite{Lester:2014aa},
which only act to alter the quantitative aspects of transport and
dispersion. As the network model considered herein is the simplest
representation of porous media with non-trivial pore-scale topology
(which is common to all porous media), we contend that the qualitative
transport dynamics of this model are universal. In
Section~\ref{sec:real_media} we demonstrate how these results can be
extended to real porous architectures as characterised by
e.g. micro-CT imaging.

The fundamental building block of the random network model is the
symmetric pore branch element $\Omega$
(Figure~\ref{fig:pore_branch}(a)), and its reflected image given by
the pore merge element (Figure~\ref{fig:pore_branch}(b)). A random
continuous 3D network can be constructed by spatial composition of
these branch and merge elements, as described in greater detail in
\ref{appendix:pore_model}.  Each random composition is unique, and a
set of compositions forms an ensemble of random networks over which
one can study the characteristics of scalar transport. The set of pore
locations in each composition defines a realization of a 3D open
porous network, and the set of all realizations of the porous network
forms a statistical ensemble which is both ergodic and stationary. If
the pore locations are uniformly and independently distributed (up to
constraints of non-overlapping pores), then the resultant distribution
of of the relative orientation of pairs of branches and merges
$\theta$ is also uniform and independent. Hence transport within the
3D open porous network follows a multidimensional Markov process in
space, forming a basis for the development of continuous-time random
walk (CTRW) models of transport, deformation and mixing.

\begin{figure}
\begin{center}
\begin{tabular}{c c}
\includegraphics[width=0.32\columnwidth]{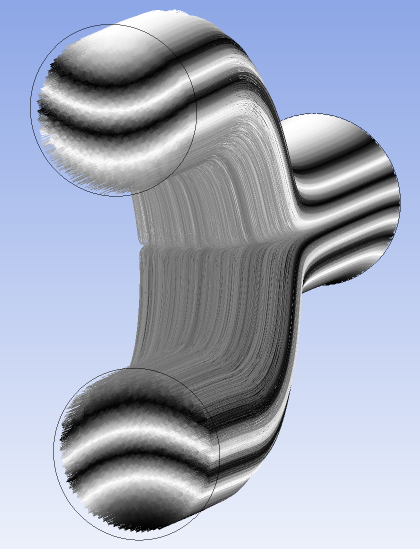}&
\includegraphics[width=0.32\columnwidth]{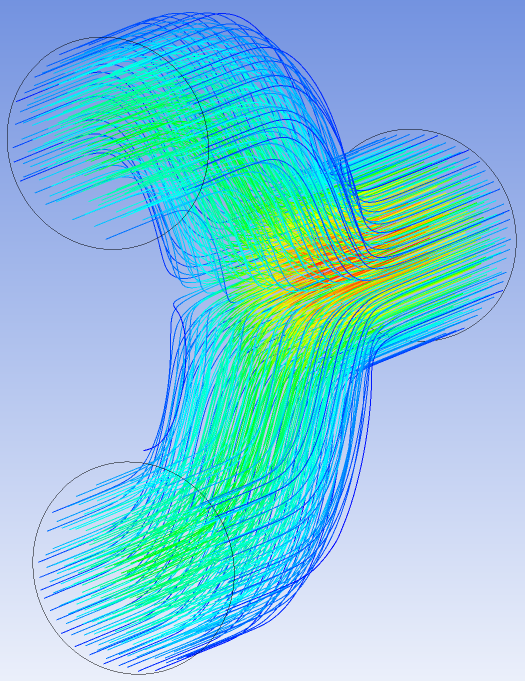}\\
(a) & (b) \\
\includegraphics[width=0.32\columnwidth]{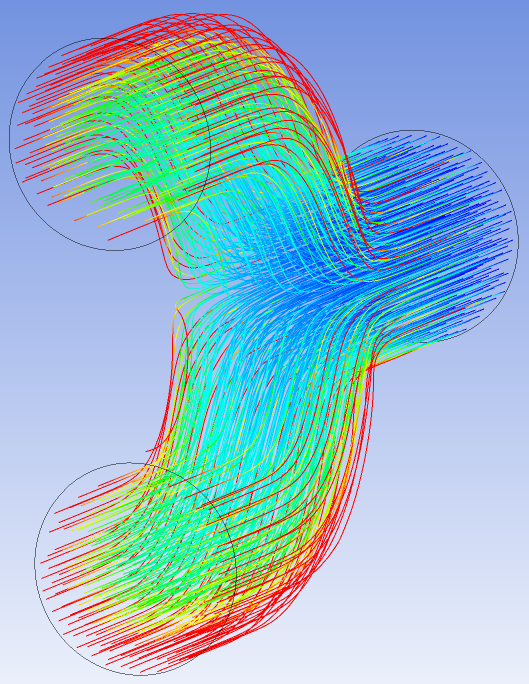}&
\includegraphics[width=0.32\columnwidth]{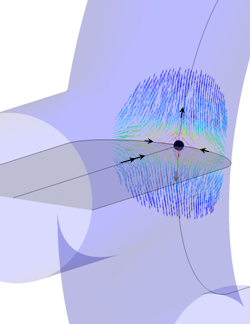}\\
(c) & (d)
\end{tabular}
\end{center}
\caption{(a) Streamlines, (b) velocity, and (c) residence time
  distributions along streamlines under Stokes flow within the pore branch element $\Omega$
  calculated from CFD calculations. (d) Saddle-type stagnation point
  (black dot), associated stable manifold $\mathcal{W}^S_{2D}$ (grey
  surface) and local skin friction vectors.}\label{fig:CFX}
\end{figure}

To model \emph{flow} within the porous network, we consider the steady
fluid velocity field $\mathbf{v}(\mathbf{x})$ within the pore branch
element $\Omega$ given by Stokes law
\begin{equation}
\mu\nabla^2\mathbf{v}+\nabla p=0,\,\,\nabla\cdot\mathbf{v}=0,
\end{equation}
forced by the pressure differential $\Delta p$ over the inlet and
outlet pores, subject to the no-slip boundary condition $\mathbf{v}|_{\partial\Omega}=0$.  The flow
field within $\Omega$ is calculated numerically to order $10^{-16}$
accuracy using the finite-volume computational fluid dynamics (CFD)
package ANSYS-CFX 14, and the resultant fluid streamlines, velocity
and residence time distributions are shown in
Figure~\ref{fig:CFX}(a)-(c). Figure~\ref{fig:CFX}(d) depicts detail of
a saddle-type stagnation point at the pore branch in $\Omega$, the
associated skin friction vector and stable manifold
$\mathcal{W}^S_{2D}$ which acts as a
minimal flux surface. Due to reversibility of Stokes flow, the
deformation, fluid velocity and skin friction fields in the pore merge
element are identical to that shown in Figure~\ref{fig:CFX} with the
velocity and skin friction field reversed, and so in this case the 2D
manifold associated with the saddle-type reattachment point is an
unstable manifold $\mathcal{W}^U_{2D}$.

\begin{figure}
\begin{centering}
\begin{tabular}{c c c}
\includegraphics[width=0.32\columnwidth]{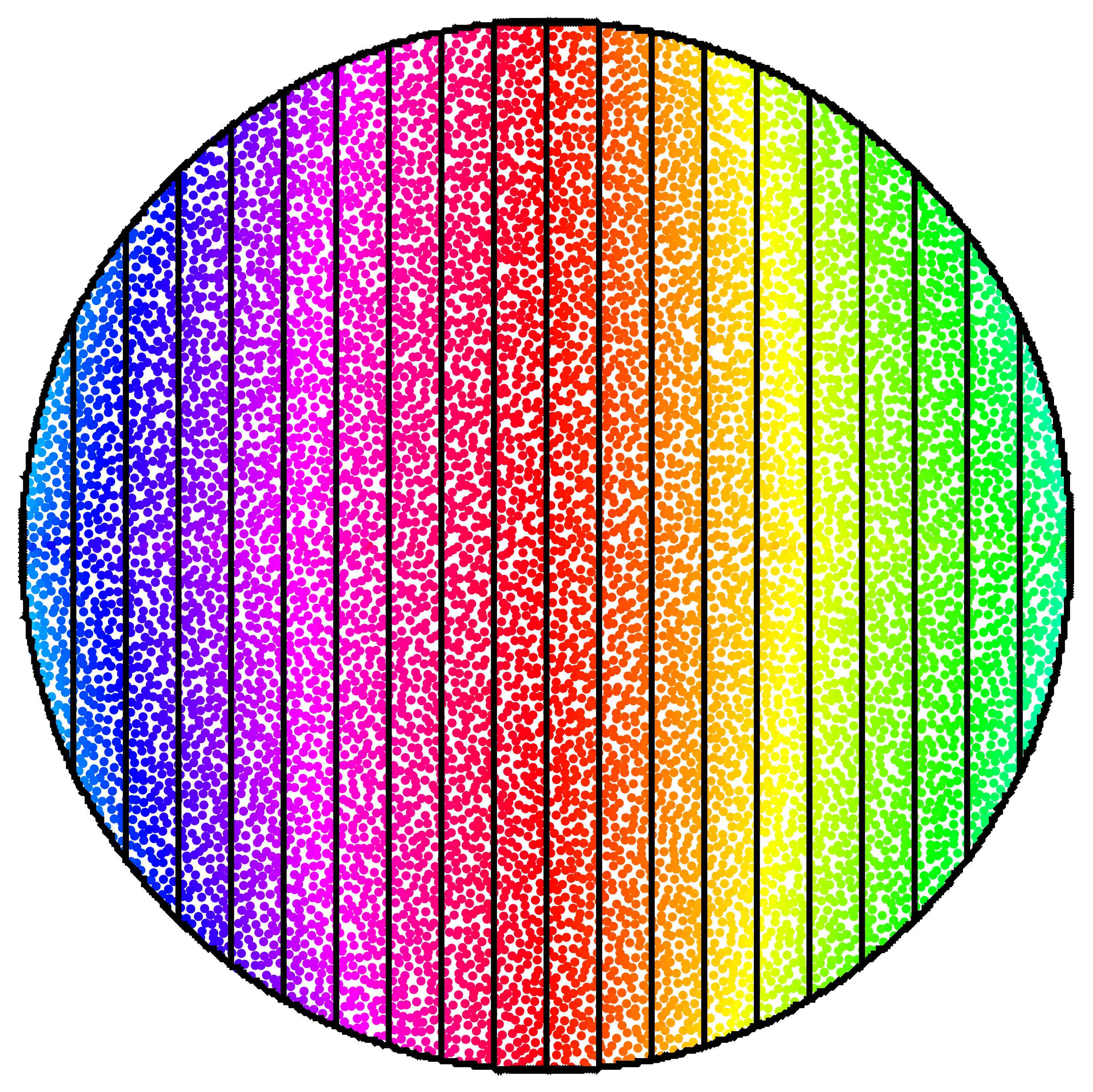}&
\includegraphics[width=0.32\columnwidth]{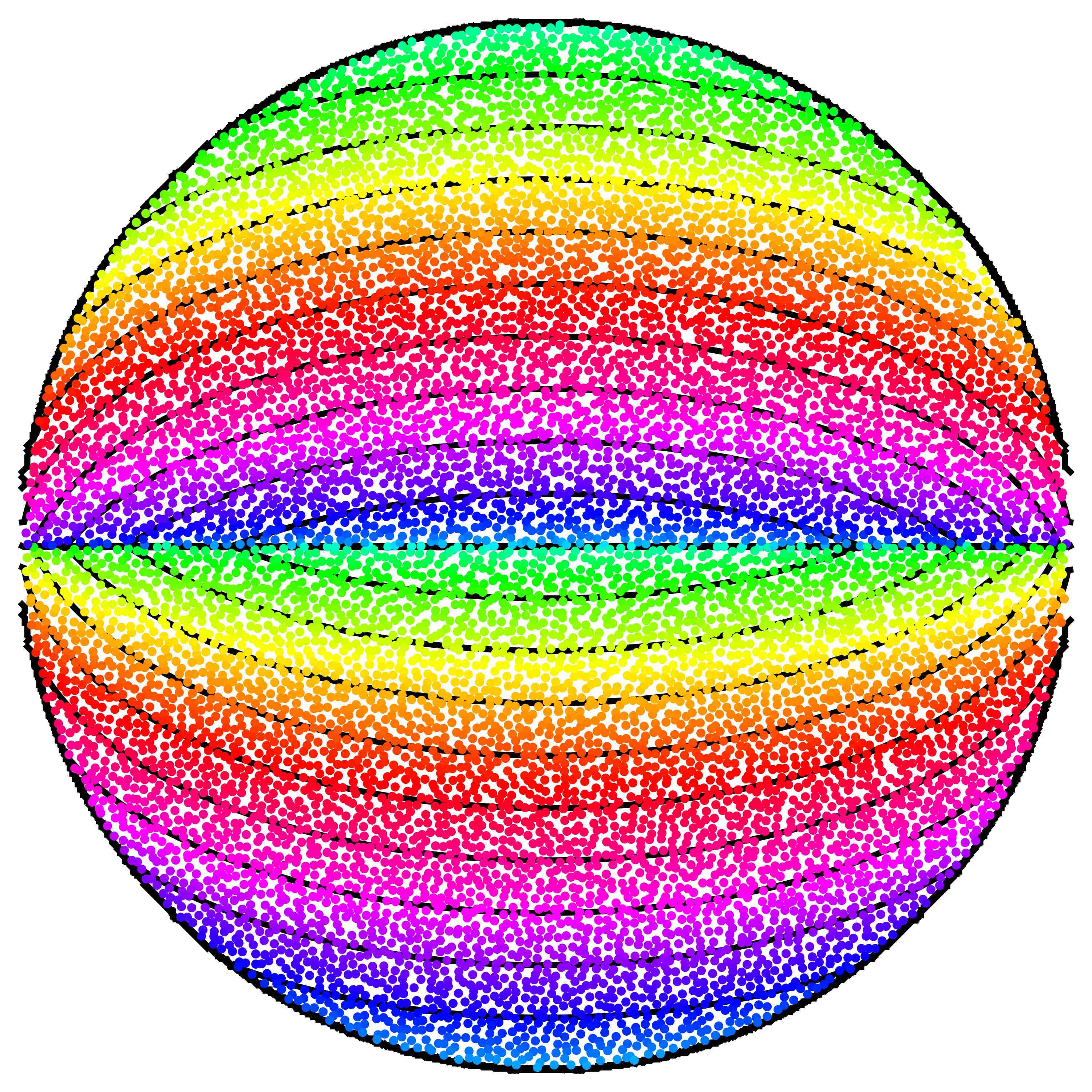}&
\includegraphics[width=0.32\columnwidth]{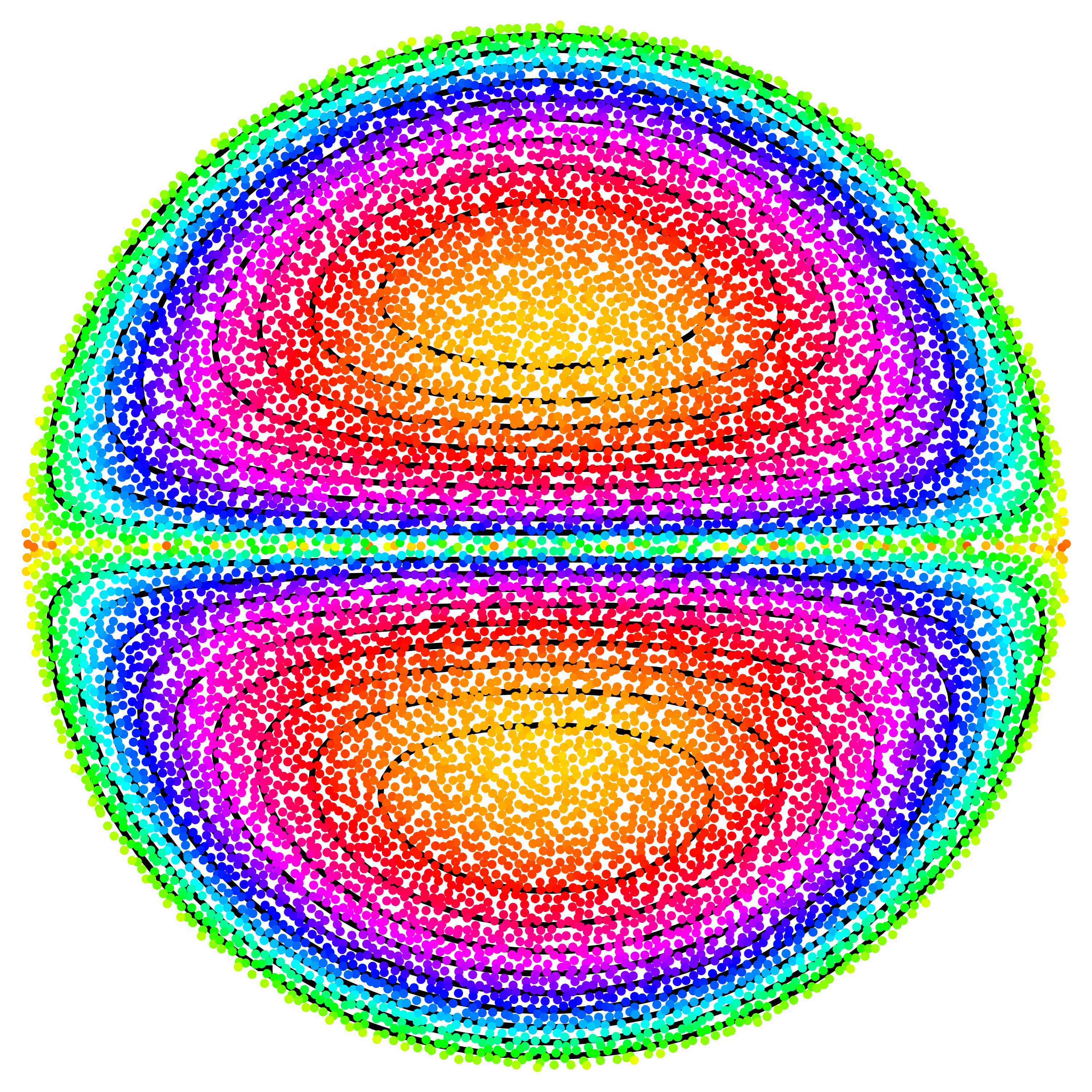}\\
(a) & (b) & (c)
\end{tabular}
\end{centering}
\caption{Comparison of distributions of (a) $x$-location, (b) $y$-location and (c) residence time in pore branch element $\Omega$ between CFD calculations (colored points) and approximate analytic maps $\mathcal{M}$ (\ref{eqn:Mmap}), $\mathcal{T}$ (\ref{eqn:Tmap}) (contour lines).}\label{fig:dist}
\end{figure}

To generate a global \emph{transport} model for the random porous
network, we consider the effect on position and time of a fluid
particle advecting through the pore branch and merge elements. We take
the results shown in Figure~\ref{fig:CFX} and define a transformation
$\mathcal{M}^\star$ which maps advection of fluid particles from
position $\{x_r,y_r\}$ at the inlet pore of $\Omega$ to the outlet
pore position $\{x_r',y_r'\}$, where $\{x_r,y_r\}$, are the scaled and
reoriented (by angle $\theta$) $x,y$ particle positions relative to
the pore boundary at the inlet and outlet, such that
$x_r^2+y_r^2=0,1$ corresponds to the pore
centre and boundary respectively. Similarly, a temporal map
$\mathcal{T}^\star:t\mapsto t'$ quantifies the residence time $t'-t$
along each particle streamline in $\Omega$. Distributions of exit
positions $x_r'$, $y_r'$ and residence times $t'-t$ are shown in
Figure~\ref{fig:dist}(a)-(c) for 100,000 streamlines within each
branch element. The exact maps $\mathcal{M}^\star$,
$\mathcal{T}^\star$ are remarkably well approximated (within relative
error $\epsilon\sim 10^{-3}$) by the simple analytic maps
$\mathcal{M}$, $\mathcal{T}$ (whose associated distributions are shown
as contours in Figure~\ref{fig:dist}(a)-(c))
\begin{align}
&\mathcal{M}:\{x_r,y_r\}\mapsto\{x_r,2y_r-\frac{|y_r|}{y_r}\sqrt{1-x_r^2}\}\label{eqn:Mmap},\\
&\mathcal{T}:t\mapsto t+\frac{1}{1-x_r^2-y_r^2}.\label{eqn:Tmap}
\end{align}
Similarly, transport in a pore merge element is quantified by the inverse spatial map $\mathcal{M}^{-1}$. The no-slip boundary condition in $\Omega$ generates the residence time distribution (\ref{eqn:Tmap}) similar to that of a Poiseuille flow, which is singular at the pore boundary, leading to unbounded residence time distributions which, as shall be shown, have a significant impact upon transport and mixing. From (\ref{eqn:Mmap}), (\ref{eqn:Tmap}), the residence time in the pore branch and merge is quantified respectively by the composite operators $\mathcal{T}\circ\mathcal{M}$, $\mathcal{T}\circ\mathcal{M}^{-1}$, respectively. For randomly oriented branch and merge elements within the porous network, advection is quantified by the composite reoriented maps respectively as
\begin{align}
\mathcal{S}(\theta)&:=R(\theta)\circ\mathcal{M}\circ R(-\theta),\label{eqn:Sbranch}\\
\mathcal{S}^{-1}(\theta)&:=R(\theta)\circ\mathcal{M}^{-1}\circ R(-\theta),\label{eqn:Smerge}
\end{align}
where $R(\theta)$ is a reorientation operator. These composite advection operators reflect symmetry breaking for $\theta\neq 0$ over a coupled pore brach and merger such that
\begin{equation}
\mathcal{S}(\theta_1)\circ\mathcal{S}^{-1}(\theta_2)\neq I\,\,\text{for}\,\,\theta:=\theta_1-\theta_2\neq 0,
\end{equation}
where $I$ is the identity operator. The approximate maps $\mathcal{S}(\theta)$, $\mathcal{T}$, capture the salient features of fluid stretching at stagnation points and hold-up at the no-slip boundaries of $\Omega$, and from Figure~\ref{fig:dist} accurately represent advection in the pore branch and merge elements. When appropriately concatenated, these maps permit highly efficient modelling of transport and mixing over a given realization of the open porous network.

Macroscopic transport in a single realization of the pore network
model is illustrated in Figure~\ref{fig:pore_network}(b), where 1000
closely spaced (equidistant over a line 1/10th of the pore diameter)
fluid particles (differentiated by colour) are advected over 100 pore
branches and mergers in the absence of molecular diffusion. Transverse
fluid stretching acts to initially spread these particles over the
first 10 pores, as illustrated by a slight broadening of the particle
trajectories, but after 10 pores the plume is split into several
different pores, a process which repeats along the network, leading to
macroscopic transverse dispersion which converges toward a Gaussian
distribution of particles, reflecting the macroscopically homogeneous
nature of the network.  This behaviour clearly depicts how pore-scale
chaotic advection impacts transverse dispersion; whilst macroscopic
transverse dispersion is chiefly governed by the geometry and topology
of the network, pore-scale chaotic advection acts to transversely
stretch fluid elements, increasing the probability of bifurcation over
a stagnation point and advection down different pores, leading to
macroscopic dispersion. Hence chaotic advection does not alter the
qualitative nature of transverse dispersion in homogeneous media, but
rather impacts the rate of dispersion. However, as shall be shown,
chaotic advection plays a significant role with respect to both
longitudinal dispersion and mixing and dilution \cite{Kitanidis:1994}
within a plume.

Figure~\ref{fig:dyetrace} shows the evolution of a dye trace simulation (consisting of the advection of a large number of diffusionless particles) propagated by the composite maps following a single pore through a realization of the random network (where the half blue/half red distribution of points is initiated across all pores), illustrating the formation of exponentially elongated material striations as the fluid approaches the well-mixed state. This behaviour is ubiquitous across all realizations of the random network, and non-mixing regions can only arise in ordered networks. Whilst there is no clear evidence of the folding of fluid elements in the $xy$-cross sections in Figure~\ref{fig:dyetrace}, folding occurs in the $z$-direction as fluid elements flow over stagnation points, and so manifests as discontinuities when projected into the $xy$-plane. The spatial and temporal maps $\mathcal{S}$, $\mathcal{T}$ implicitly capture this behaviour through the advection and residence time distributions.

\begin{figure}
\begin{centering}
\begin{tabular}{c c c}
\includegraphics[width=0.3\columnwidth]{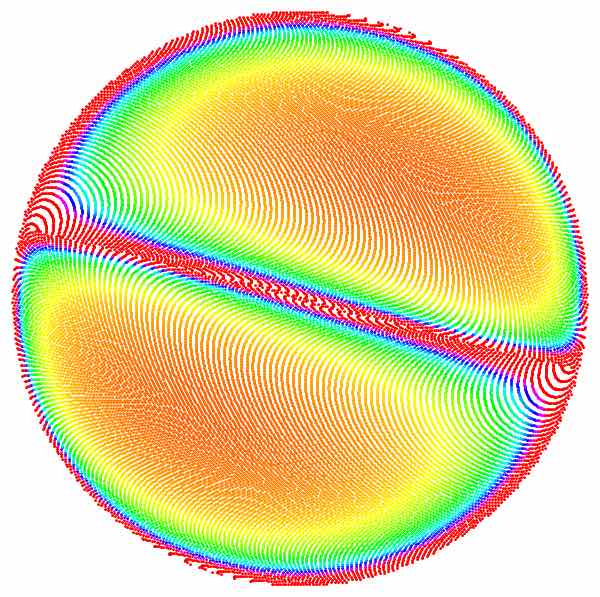}&
\includegraphics[width=0.3\columnwidth]{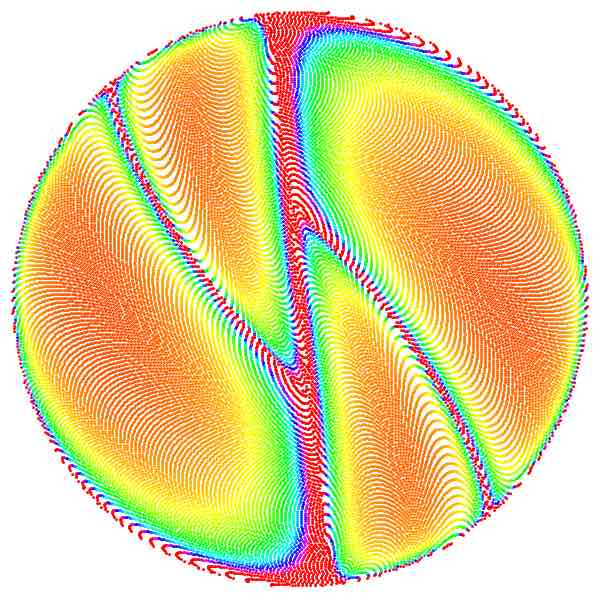}&
\includegraphics[width=0.3\columnwidth]{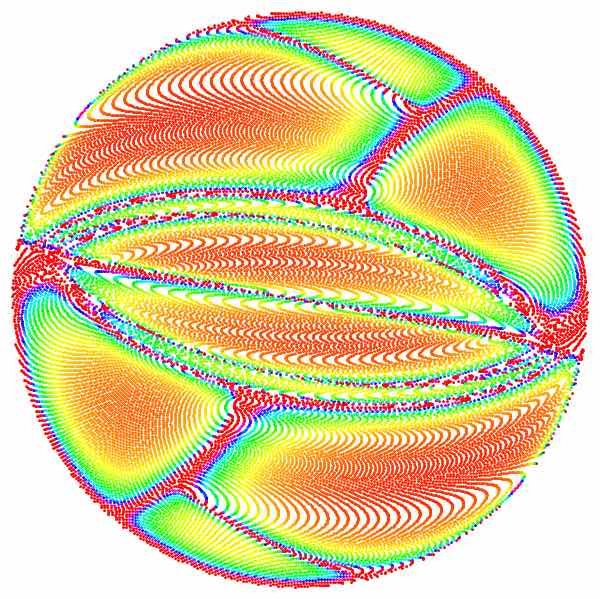}\\
$n$=1 & $n$=2 & $n$=5 \\
\includegraphics[width=0.3\columnwidth]{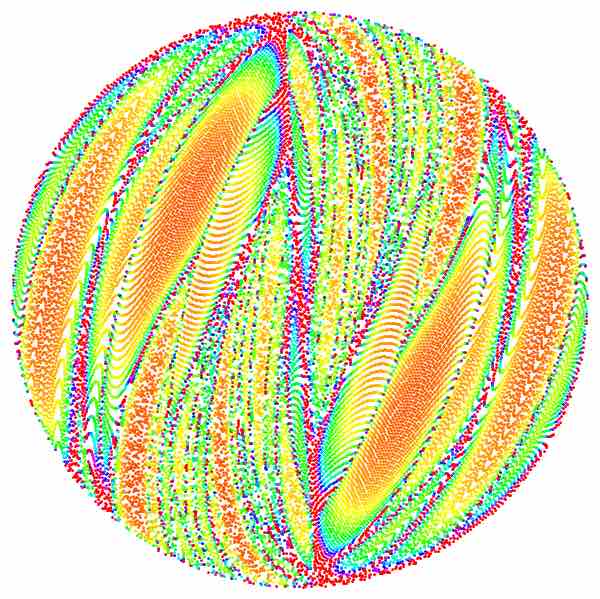}&
\includegraphics[width=0.3\columnwidth]{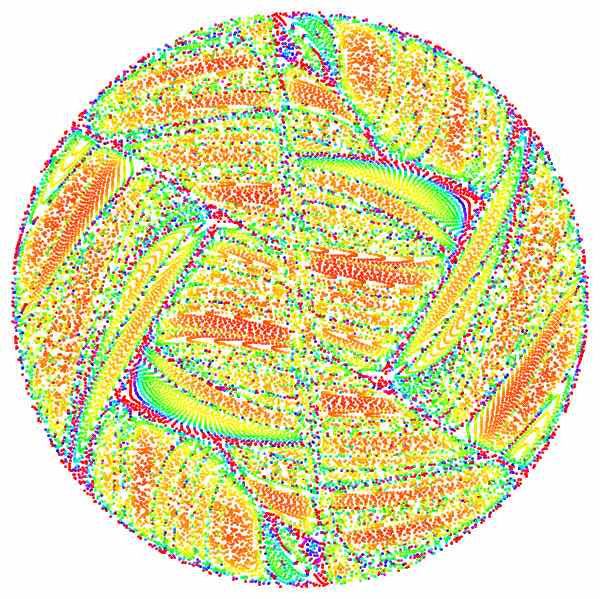}&
\includegraphics[width=0.3\columnwidth]{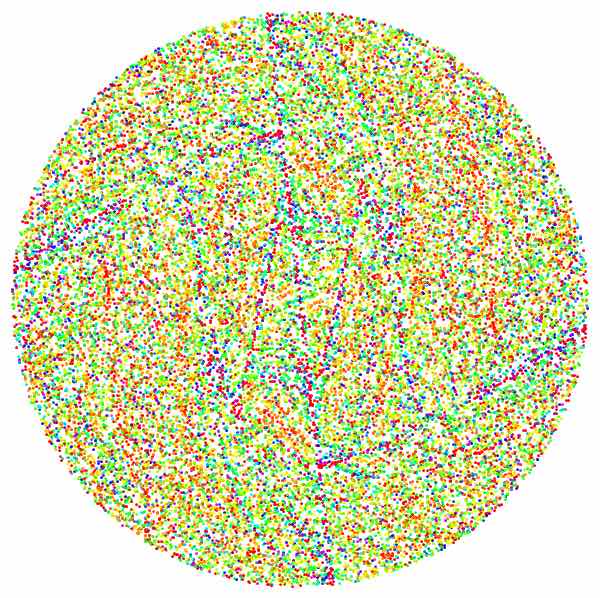}\\
$n$=10 & $n$=20 & $n$=50
\end{tabular}
\end{centering}
\caption{Calculated residence time distributions (pink = large, orange=small) with increasing number of pore branch-merge transitions $n$.}\label{fig:RTD}
\end{figure}

Evolution of the residence time distribution for these points is shown
in Figure~\ref{fig:RTD}, where the large residence times of the
temporal map $\mathcal{T}$ associated with the no-slip boundary
condition are clearly illustrated for small values of the longitudinal
pore number $n$. The spatial distribution of residence times become
increasing striated as particles are propagated through the network,
and so chaotic advection acts to ``mix'' these residence times, as
shown for large $n$, i.e. the residence time
  distribution in pores becomes more and more peaked and uniform,
  albeit retaining a long tail in the distribution.  Whilst, in the
absence of diffusion, particle trajectories and hence residence times
are not random, the space-filling nature of chaotic orbits renders
these trajectories to be ergodic, and so may be conceptualised as a
random process with decaying correlation along streamlines, where this
decay rate is directly related to the Lyapunov exponent
$\lambda_\infty$. As shall be shown, the impact of such ``mixing'' of
residence times has a significant impact upon macroscopic longitudinal
dispersion.

\section{Pore-scale mixing due to chaotic advection-diffusion}\label{sec:pore_mixing}

The results above illustrate the mechanisms which generate chaotic
advection in 3D porous media and the consequent impacts upon fluid
stretching and material distributions. To quantify the impact of
chaotic advection upon pore-scale mixing of a tracer solute, we summarise the results of Lester et
al.~\cite{Lester:2016aa} regarding the interplay of such fluid deformation with molecular
diffusion. For simplicity of exposition we consider the
steady concentration field $c(\mathbf{x})$ which arises from a
continuous source at the inlet plane $z=0$ which is uniformly injected
across all inlet pores as a line source with Gaussian concentration
profile (given by (\ref{eqn:gauss_pulse})). Whilst this inlet condition is not representative
of any physically important situation, this condition renders the
macroscopic concentration distribution uniform and so more clearly
illustrates the mixing and dilution dynamics. Results derived for this
specific condition are then generalised to an arbitrary inhomogeneous macroscopic
concentration distribution at the end of this section.

The concentration distribution within the pore space is described by the steady 3D advection-diffusion equation (ADE)
\begin{equation}
-\mathbf{v}(\mathbf{x})\cdot\nabla c(\mathbf{x})+D_m\nabla^2 c(\mathbf{x})=0,\label{eqn:3D_ADE}
\end{equation}
subject to the zero-flux boundary condition
$\mathbf{n}\cdot\nabla c|_{\partial \mathcal{D}_f} =0$, where
$\mathbf{n}$ is the outward normal vector from the fluid domain
$\mathcal{D}_f$, and where $D_m$ is the molecular diffusivity. Following~\cite{Lester:2016aa} we ignore longitudinal diffusion for the continuous injection protocol and thus pose the steady 3D ADE
(\ref{eqn:3D_ADE}) as a transient 2D ADE in the $xy$-plane as
\begin{equation}
\frac{\partial c}{\partial t}=v_z\frac{\partial c}{\partial z}=-\mathbf{v}_\bot(x,y,t)\cdot\nabla_{\bot}c(x,y,t)+D_m\nabla^2_\bot c(x,y,t),\label{eqn:sep_ADE}
\end{equation}
where $v_z$ is the $z$-component of $\mathbf{v}(\mathbf{x})$, $\mathbf{v}_\bot$ and $\nabla_\bot$ denote the $xy$-components of velocity and gradient operator respectively. The $z$-coordinate is also parameterised in terms of the Lagrangian travel time $t=t(z)$ along a given fluid particle trajectory as
\begin{equation}
t(z)=\int_0^z dz^\prime \frac{1}{v_z(z^\prime)}.\label{eqn:L_time}
\end{equation}
Whilst the travel time $t(z_0)$ for fixed $z=z_0$ also varies over the  $xy$-plane, (\ref{eqn:sep_ADE}) provides a convenient basis for solution of the concentration field based upon a CTRW for the local advection time and deformation history of fluid elements.

For highly striated, lamellar material distributions (such as those generated by chaotic advection) the ADE simplifies to a 1D transport equation~\cite{Ranz:1979aa,Duplat:2008aa,Duplat:2010aa,Le-Borgne:2013aa} for an individual lamella \textit{l} in terms of the material coordinates $\{\chi,\eta\}$ where $\eta$ is the coordinate transverse to the elongated lamella \textit{l} and $\chi$ is the coordinate along the lamella. Due to fluid stretching, concentration gradients along the lamella ($\chi$) are exponentially weakened,  whilst gradients transverse to the lamellae ($\eta$) are exponentially amplified, and so (\ref{eqn:sep_ADE}) further simplifies to
\begin{equation}
\frac{\partial c}{\partial t}-\gamma(t)\eta\frac{\partial c}{\partial\eta}=D_m\frac{\partial^2c}{\partial\eta^2},\label{eqn:1D}
\end{equation}
where the temporal stretching rate of a lamella, $\gamma(t)$, is given by the elongation rate
\begin{equation}
\gamma(t)=\frac{d\ln \rho(t)}{dt},
\end{equation}
and $\rho(t)=\ell(t)/\ell(0)$ is the local deformation of a material line relative to its original length $\ell(0)$. The ensemble average of the stretching rate $\gamma(t)$ converges toward a constant for exponential stretching and scales as $1/t$ for algebraic fluid stretching, and so the fluid deformation rate directly impacts the evolution of the lamellae. For a Gaussian initial concentration profile in each pore of the form
\begin{equation}
c(\eta,0)=\frac{c_0}{\sqrt{2\pi\sigma_0^2}}\exp\left(-\frac{\eta^2}{2\sigma_0^2}\right),\label{eqn:gauss_pulse}
\end{equation}
solutions to the lamellar ADE (\ref{eqn:1D}) are given by
\begin{equation}
c(\eta,t)=\frac{c_0}{\sqrt{2\pi(\sigma_0^2+2D_m\tau(t))}}\exp\left(-\frac{\eta^2\rho(t)^2}{2\sigma_0^2+4D_m\tau(t)}\right),\label{eqn:1DADE}
\end{equation}
where the maximum concentration across the lamella is $c_m(t)=c(0,t)$ and the operational time $\tau(t)$ which encodes the entire stretching history is defined as
\begin{align}
&\tau(t)=\int_0^t\rho(t^\prime)^2dt^\prime.
\end{align}
Hence the action of fluid stretching as quantified by $\rho(t)$ acts to both narrow the striation due to fluid stretching, but also increase the profile variance as $\sigma^{\star2}(t)=\sigma_0^2(t)+2D_m\tau(t)$. Note that (\ref{eqn:1DADE}) quantifies the transverse concentration profile at a single point along the lamellar backbone (denote by the $\chi$-coordinate), and so the concentration field over a single is described as $c(\chi,\eta,t)$. The entire concentration field is then comprised of the set of lamellae undergoing distortions as they migrate through the pore-space. 

To develop analytic solutions for the evolving concentration field,
Lester et al.~\cite{Lester:2015aa} develop a stretching CTRW in the open porous network which
describes evolution of fluid deformation with pore number $n$ as a two-step process (corresponding to
alternating pore branches and mergers) for the advection time $t$ and length $\ell$ (in the $\chi$-direction) of an
element of a material strip (lamella) in the 2D plane transverse to
the mean flow direction as
\begin{align}
&\ell_{n+1}=\ell_n \rho_s(\varphi_n),\quad\quad t_{n+1}=t_n+\Delta t_n,\label{eqn:rhos}\\
&\ell_{n+2}=\ell_{n+1} \rho_c(\varphi_{n+1}),\quad\quad t_{n+2}=t_{n+1}+\Delta t_n,\label{eqn:rhoc}
\end{align}
where $n$ is the pore number and $\Delta t_n$ is the advection time
between pores. $\rho_s(\theta)$, $\rho_c(\theta)$ are the relative
elongations due to stretching and compression, respectively, of $\ell$
over a pore branch and merger, and $\varphi$ is the orientation of the
material strip with respect to the pore branch or merger which, due to randomness of the pore network, is distributed uniformly over $[-\pi,\pi]$. Note that $\theta$ represents the angle between a pore branch and merger, whereas $\varphi$ represents the angle between the material strip and the local branch or merger. Closed expressions for the relative elongations
$\rho_s$, $\rho_c$ are derived directly from the advection maps
$\mathcal{S}(\theta)$, $\mathcal{S}^{-1}(\theta)$ as
\begin{align}
&\rho_s(\varphi)=\sqrt{1+3\cos^2\varphi},\label{eqn:rhosdist}\\
&\rho_c(\varphi)=\frac{1}{2}\sqrt{4-3\cos^2\varphi}.\label{eqn:rhocdist}
\end{align}
Similarly, the distribution of advection time between pores
$\Delta t_n$ determined from the temporal map $\mathcal{T}$ is distributed as a Pareto variable
\begin{align}
\psi(\Delta t)=H(\Delta t-\Delta t_a)\frac{1}{\Delta t_a}\left(\frac{\Delta t}{\Delta t_a}\right)^{-2},\label{eqn:Pareto}
\end{align}
where $H$ is the Heaviside step function and $\Delta t_a$ is the
minimum waiting time in the pore.  As discussed above, the ergodicity
of chaotic orbits permits the space-filling nature of fluid particle
trajectories to be modelled as the stochastic process defined by
equations~(\ref{eqn:rhos}) and (\ref{eqn:rhoc}).  This stretching CTRW
captures the essential features of exponential fluid stretching and
arbitrarily long waiting times as the fluid particles propagate
through the porous medium.  As a result of the Central Limit Theorem,
the total fluid deformation $\rho=\rho_s\rho_c$ over a coupled pore
branch/merger converges with pore number $n$ toward a lognormal
distribution
\begin{equation}
\hat{p}_\rho(\rho)=\frac{1}{\rho\sqrt{\pi n\sigma^2}}\exp\left[-\frac{(\ln\rho-n\lambda_\infty/2)^2}{n\sigma^2}\right],\label{eqn:rhodist}
\end{equation}
where the asymptotic average stretching rate
$\lambda_\infty\approx 0.1178$ and variance
$\sigma^2\approx 0.11436$ over the pore network are derived from (\ref{eqn:rhosdist}),
(\ref{eqn:rhocdist}).  This result allows the
evolution of the total deformation $\rho$ given by the two-step CTRW (\ref{eqn:rhos}) (\ref{eqn:rhoc}) to be expressed in terms of of a
one-step CTRW for the relative stretch $\rho_n$ of the line element as  
\begin{equation}
l_{n+1}=l_n \rho_n,\quad\quad t_{n+1}=t_n+\Delta t_n,\label{eqn:1stepCTRW}
\end{equation}
which may be solved
analytically~\cite{Lester:2015aa}.  To solve diffusive mixing in the
pore network, it is necessary to also solve the operational time
$\tau(t)$ in (\ref{eqn:1DADE}), which is well-approximated by the
expression
\begin{equation}
\tau_n:=\tau(t_n)\approx\frac{t_n \rho_n^2}{n\Lambda_\infty},\label{eqn:tau_history}
\end{equation}
where $\Lambda_\infty=\lambda_\infty+\sigma^2$. These results
completely quantify evolution of the lamellar mixing model
(\ref{eqn:1DADE}) up to the coalescence regime, where previously
non-interacting lamellae merge into a continuous mixing plume, and the
1D approximation governing (\ref{eqn:1DADE}) breaks down. Methods are
available \citep{Duplat:2008aa,Le-Borgne:2013aa} to predict evolution
of the scalar concentration field beyond this \emph{coalescence
  limit}, and whilst these are beyond the scope of this paper, the
rate of mixing is still strongly dependent upon the fluid stretching
dynamics, hence chaotic advection augments fluid mixing in this regime
also.

As shown by Lester et al.~\cite{Lester:2015aa} and Le Borgne et al.~\cite{Le-Borgne:2015aa}, this CTRW formulation leads to asymptotic estimates of the decline in average maximum concentration of the tracer plume as the pore number $n$ increases. 
In 2D porous networks, the asymptotic maximum concentration $c_m$ decays \textit{algebraically} with $n$ as 
\begin{equation}
\langle c_{m,2D}(n)\rangle\propto n^{-\alpha-\frac{\beta+1}{2}},\label{eqn:cm2D}
\end{equation}
where $\alpha \geqslant 1/2$ and $\beta \leqslant 2$ are characteristic exponents of the underlying velocity distribution~\cite{Dentz:2015aa,Dentz:2015ab}. However, in 3D porous networks the asymptotic maximum concentration declines \textit{exponentially} with $n$ as
\begin{equation}
\langle c_{m,3D}(n)\rangle\propto \frac{1}{\sqrt{\ln(n)}} \exp\left[-n \frac{2 \lambda_\infty+\sigma^{2}}{4}\right].\label{eqn:cm3D}
\end{equation}
In terms of ergodic theory 2D steady porous media flows
  intrinsically produce, at best, weak mixing, and 3D steady porous
  media flows intrinsically produce strong mixing.

The fundamental difference between mixing in 2D and 3D porous media is
clear.  The Poincar\'{e}-Bendixson theorem \cite{Teschl:2012aa}
excludes chaotic advection in 2D steady flows but permits it in 3D
steady flows.  This phenomenon must occur and even be ubiquitous in
topologically complex porous media.  The major observable effect is a
transition from algebraic fluid stretching in two dimensions to
exponential stretching in three dimensions that leads to exponential
dilution at the pore scale in real systems as a function of
longitudinal distance(the pore number $n$) from a point source.

This behaviour is also observed in the rate of scalar dissipation (quantified as variance decay), where Lester et al~\cite{Lester:2015aa} show that concentration variance across a pore of area $A$ varies with pore number $n$ directly as
\begin{equation}
\sigma_c^2(n):=\frac{1}{A}\int_A(c(\eta,t_n)-\overline{c(n)})^2dA=\overline{c(n)}^2\left(\frac{\langle c_m(n)\rangle}{\sqrt{2}\,\overline{c(n)}}-1\right),
\end{equation}
where $\overline{c(n)}$ is the average concentration across the pore and $\langle c_m(n)\rangle$ is the average concentration over the concentration support area, defined as the area corresponding to $c$ greater than an cut-off concentration $\epsilon$ which is small but finite~\cite{Lester:2015aa}.  For moderate $n$, where
$\langle c_m(n)\rangle\gg \overline{c(n)}$, the scalar variance scales
as $\langle c_m(n)\rangle$, and so the rate of fluid stretching
directly impacts the rate of scalar dissipation.  Again, the chaotic
advection inherent to 3D porous media imparts dissipation which scales exponentially with pore number $n$, whereas algebraic dissipation obtains in 2D porous media.

These results are generalised in~\cite{Lester:2015aa} to the case of an arbitrary heterogeneous macroscopic
concentration distribution, which may be defined in terms of the total length
$l_p(\mathbf{x})$ of lamellae per pore local to $\mathbf{x}$.
For such a macroscopic distribution, the concentration mean and variance within the plume then vary as
\begin{align}
&\overline{c_p(\mathbf{x})}=\overline{c(n)}\frac{l_p(\mathbf{x})}{l(n)},\\
&\sigma_{c_p}^2=\overline{c_p(\mathbf{x})}^2\left(\frac{\langle c_m(n)\rangle}{\sqrt{2}\overline{c_p(\mathbf{x})}}-1\right),
\end{align}
and again the rate of scalar dissipation within a heterogeneous plume is also governed by the rate of fluid stretching via $\langle c_m(n)\rangle$. Hence chaotic advection acts to significantly alter the dynamics of mixing and dilution at both the pore and macro scales.

\section{Impact of chaotic advection upon longitudinal dispersion}\label{sec:long_disperse}

In the previous Section we have considered the impact of pore-scale
chaotic advection upon mixing and dilution. As illustrated in
Figure~\ref{fig:RTD}, these kinematics also significantly impact
evolution of the residence time distribution and hence longitudinal
dispersion, as is well established over several
studies~\cite{ Mezic:1999aa,Jones:1994aa,Metcalfe:2006aa,Lester:2015ab,Lester:2016ab} of
chaotic duct flows. Pore-scale chaotic advection imparts particle
trajectories which, whilst deterministic, are ergodic due to the
space-filling nature of the chaotic orbits. Hence diffusion-less fluid
trajectories sample all of the velocities across the pore space as
they are advected downstream, retarding longitudinal dispersion in a
manner similar to Taylor-Aris dispersion. Specifically, in the absence
of diffusion, pore-scale chaotic advection acts to retard growth of
longitudinal variance $\sigma_L^2(t)$ (related to longitudinal
dispersion as $D_L(t ) = \sigma_L^2(t)/2t)$ from ballistic growth
$\sigma_L^2(t)\sim t^2$ due to the no-slip boundary condition to
super-diffusive growth $\sigma_L^2(t)\sim t^\alpha$, $1 < \alpha <
2$.
While analogous to Taylor-Aris dispersion in the presence of molecular
diffusion, this mechanism is distinctly different in that chaotic
advection is hydrodynamic (geometric) in origin and deterministic and
so cannot be modelled as a simple Fickian diffusion process. These
dynamics also persist in the presence of molecular diffusion in that
chaotic advection provides a mechanism to significantly accelerate
transverse mixing, which in turn further retards longitudinal
dispersion~\cite{Lester:2015ab,Lester:2016ab}.

In this section we first quantify the impacts of pore-scale chaotic
advection upon longitudinal dispersion in the absence of molecular
diffusion, prior to discussion of extension of these results to
diffusive tracers. Whilst chaotic orbits are not random, these fluid
trajectories exhibit decaying correlations (the decay rate of which is
related to the Lyapunov exponent $\lambda_\infty$), and so follow a
Markov process when appropriately spatially discretised. This property
allows diffusion-less tracer advection in the 3D open porous network
to be modelled as a CTRW, which captures the interactions between
tracer holdup due to the no-slip boundary condition and chaotic
advection due to fluid stretching around stagnation points of the
flow.

\begin{figure}
\begin{centering}
\begin{tabular}{c}
\includegraphics[width=0.9\columnwidth]{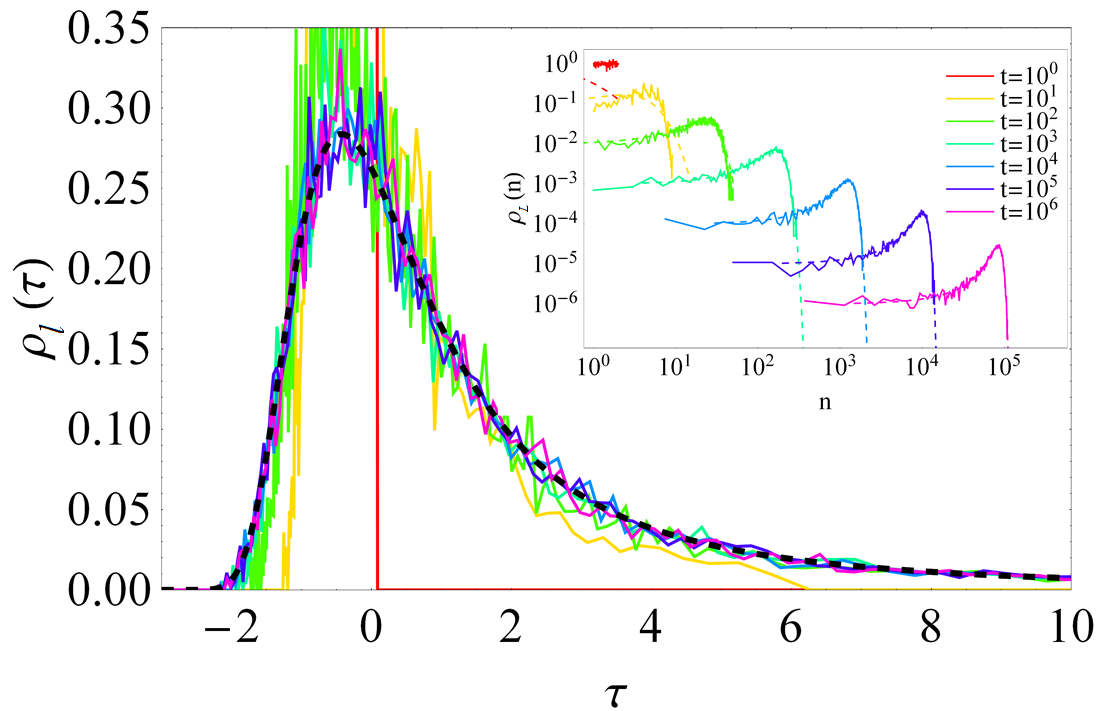} \\
(a) \\
\includegraphics[width=0.9\columnwidth]{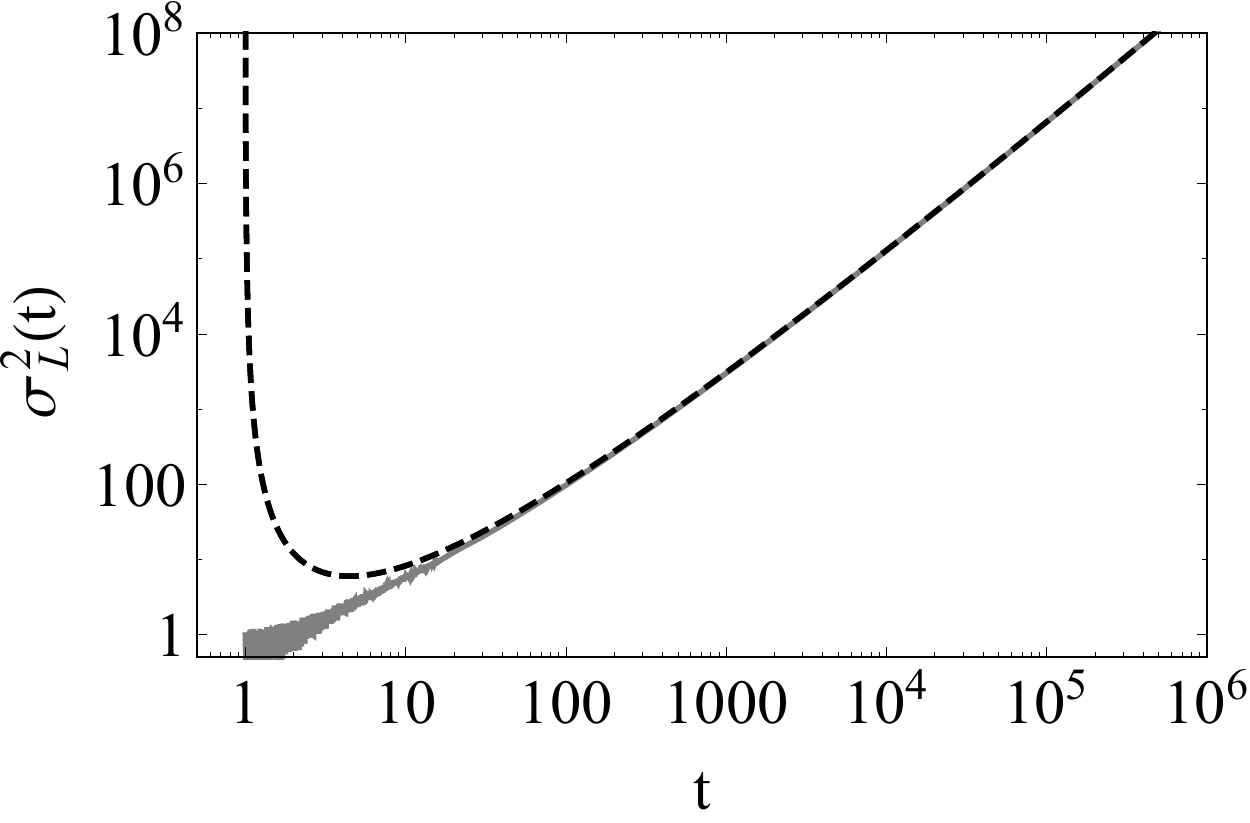}\\
(b)
\end{tabular}
\end{centering}
\caption{(a) (Color online) Convergence of residence time distribution
  for $10^5$ orbits in a single realization of the random network upon
  shifting and scaling to the Landau distribution $\rho_l(t)$. Inset:
  comparison between analytic (dashed) and numerical (solid) axial
  distribution $\rho_L(n)$. (b) Longitudinal dispersion
  $\sigma_L^2(t)\sim t^2/(\ln t)^3$ as predicted by asymptotic
  estimate in (\ref{eqn:asymptote}) (dashed, black) and pore network
  simulations (solid, grey).}\label{fig:long_dist}
\end{figure}

In Lester et al.~\cite{Lester:2014aa} we formalized this CTRW in terms
of the displacements $\Delta\mathbf{x}$ and transition times
$\Delta t$ over each pore branch and merge element of the flow (shown
in Figure~\ref{fig:pore_branch}) which are populated from the CFD
computations outlined in Section~\ref{sec:network_model}. If we
consider the evolution of a fluid particle propagating in the mean
flow direction $z$, then in the absence of diffusion the longitudinal
position $z$ and residence time $t$ of fluid particle evolves via the
CTRW as
\begin{align}
t_{n+1}&=t_{n}+\Delta t_n,\label{eqn:CTRW_time}\\
z_{n+1}&=z_{n}+\Delta z,\label{eqn:CTRW_space}
\end{align}
where the temporal increment $\Delta t$ is distributed via the PDF $\psi(\Delta t)$ (\ref{eqn:Pareto}), and for the model network the spatial increment $\Delta z$ is constant (corresponding to the distance between mapping planes in Figure~\ref{fig:pore_network}). Solution of this CTRW~\cite{Lester:2014aa} shows that the axial distribution of fluid particles $p_L(n,t)$ (expressed in terms of longitudinal pore number $n$) rapidly converges to a Landau distribution $p_l(\tau)$
\begin{align}
p_L(n,t)&=\frac{2}{\pi}\left(\frac{1}{n}+\frac{t}{n^2}\right)p_l\left[\frac{2t-2n\left(\ln(n+1)-\gamma^\star-\ln\frac{2}{\pi}\right)}{n\pi}\right],\\
p_l(\tau)&=\frac{1}{\pi}\int_0^\infty \exp(-\zeta\ln\zeta-\tau\zeta)\sin(\pi\zeta)d\zeta,
\end{align}
where $\gamma^\star$ is Euler's constant. As shown in
Figure~\ref{fig:long_dist}(a), this prediction agrees well with direct
numerical simulations via the advection and temporal maps
$\mathcal{S}$, $\mathcal{T}$ for the open porous network. The
asymptotic longitudinal variance of the resulting distribution is also
estimated as
\begin{equation}
\sigma_L^2(t)\sim\frac{t^2}{(\ln t)^3},\label{eqn:asymptote}
\end{equation}
which agrees well with predictions from the 3D pore network model at
long times ($t\gtrsim 10^3$) (Figure~\ref{fig:long_dist}(b)). Hence
the impact of chaotic advection in the 3D model network is to retard
longitudinal dispersion from a ballistic phenomenon
$\sigma_L^2(t)\sim t^2$ to super-diffusive anomalous transport
$\sigma_L^2(t)\sim t^2/(\ln t)^3$.

For real homogeneous porous media the topology and geometry of the
pore space is more complex than is catered for in the idealised 3D
network that we use here for illustration.  This difference would lead
to alterations in the transverse stretching rates $\lambda_\infty$,
and spatial $\rho_{\Delta x}$ and temporal $\psi(\Delta t)$ increments
in the CTRW model. It is important to note that the Lyapunov exponent
$\lambda_\infty$ does not directly control longitudinal plume
dispersion; rather $\lambda_\infty$ ensures ergodicity of fluid orbits
which renders the assumptions of the CTRW model valid, and so only the
spatial and temporal increments control longitudinal dispersion. The
asymptotic scaling result (\ref{eqn:asymptote}) holds for all
$\rho_{\Delta x}$ under the Poiseuille (Pareto) time increment
(\ref{eqn:Pareto}). In general, the CTRW longitudinal dispersion for
any porous medium can be derived, given $\psi(\Delta t)$,
$\rho_{\Delta x}(\Delta x)$.  As discussed by Lester et al.~\cite{Lester:2014aa},
the no-slip boundary condition generates unbounded temporal increments
$\psi(\Delta t)$ and, as a consequence, for any spatial increment
distribution $\rho_{\Delta x}(\Delta x)$, the CTRW dynamics lead to
pre-asymptotic and asymptotic transport which is universally anomalous
(super-linear but sub-ballistic). 

These results may also be extended to the case of diffusive tracers, where pore-scale chaotic advection acts to significantly accelerate scalar mixing, essentially enhancing the Taylor-Aris mechanism for molecular diffusion, and so further retarding longitudinal dispersion. In the presence of molecular diffusion, the asymptotic longitudinal dispersion dynamics are rendered Fickian, and so may be quantified solely in terms of an asymptotic dispersion coefficient $D_{L,\infty}$. These dynamics are most clearly illustrated via consideration of the mathematical connection between transverse mixing and RTD evolution in boundary-dominated flows~\cite{Lester:2015ab}. In general, the rate of transverse mixing scales with Pecl\'{e}t number $Pe$ as $Pe^\alpha$, where $\alpha\approx1/2$ for chaotic advection. The relationship between longitudinal dispersion and traverse mixing then yields~\cite{Lester:2015ab,Lester:2016ab}
\begin{equation}
D_{L,\infty}=D_m\left(1+\kappa_0Pe^{2-\alpha}D_m\right),\label{eqn:long_dispersion}
\end{equation}
In general, accelerated transverse mixing directly retards asymptotic
longitudinal dispersion. Whilst diffusive longitudinal dispersion is
yet to be couched in terms of stochastic models, the physical
mechanisms are generic to all porous media, time scales and length
scales, and have profound effects upon dispersion at the
macroscale. Hence, pore-scale chaotic advection significantly augments
dispersion and mixing at the macroscale in all porous media.

\section{Application to Non-Ideal Porous Architectures} \label{sec:real_media}

The pore-scale advection dynamics and implications for dispersion and
mixing analysed in
sections~\ref{sec:network_model}-\ref{sec:long_disperse} are based
upon idealised models of random porous media and leading-order
analysis of the deformation dynamics and associated mixing
phenomena. Whilst quantitative, these idealised models act to
illustrate the governing mechanisms and resultant scalings for mixing
and dispersion. To develop more general theories of macroscopic
transport which capture the effects of more realistic pore-scale
phenomena, it it necessary to extend these results to non-ideal media,
such as pore space architectures given by micro-CT imaging studies.

Such an approach not only facilitates the development of stochastic
models of mixing and dispersion which honour the pore-scale advection
dynamics but also provide important data with respect to linking
measures of pore-scale geometry (such as topology, tortuosity,
heterogeneity) of real media to stochastic mixing kernels. This
approach both identifies the pore-scale features which govern
transport and mixing and paves the way for development of stochastic
models which are couched directly in terms of medium properties. By
utilising recent developments in lamellar mixing
models~\cite{Le-Borgne:2013aa,Le-Borgne:2015aa} for porous media which
use deformation dynamics as inputs, it is possible to predict
evolution of the full concentration PDF within the plume, which is
important for upscaling of mixing and nonlinear chemical reactions.

Although exponential fluid stretching due to pore-scale chaotic advection dominates over algebraic deformation due to shear in the asymptotic limit, convergence to this regime may be slow, and so shear deformation can also play an important role in the pre-asymptotic regime. Moreover, fluid deformation (strain) is a tensorial, rather than scalar, quantity, and so to capture all of the augmented transport dynamics that arise from the interplay of molecular diffusion and fluid advection at the pore-scale it is necessary to properly quantify the full fluid deformation gradient tensor $\mathbf{F}$ which evolves along a streamline as
\begin{equation}
\frac{d\mathbf{F}}{dt}=[\nabla\mathbf{v}(t)]^T\cdot\mathbf{F}(t),\,\,\mathbf{F}(0)=\mathbf{I},\label{eqn:deform3D}
\end{equation}
where the advection time $t$ denotes the position $\mathbf{x}(t)$ along a streamline with initial position $\mathbf{x}(0)=\mathbf{x}_0$.

This issue is illustrated by consideration of the difference between
3D fluid mixing under continuous or pulsed injection of a line source.
Focusing solely upon the advection dynamics (i.e. ignoring diffusion),
in the continuous injection case (as considered in
Section~\ref{sec:pore_mixing}) the line source forms a continuous 2D
sheet which is stretched exponentially and folded into a 2D lamellar
structure. Deformation of this sheet is governed by longitudinal and
transverse stretching and shear, as quantified by the elements of
$\mathbf{F}$.  Conversely, transient injection of a line source (as
considered in Section~\ref{sec:long_disperse}) at the same injection
point generates a continuous 1D line which, at any point in time,
resides within the 2D sheet described above. Whilst deformation of
this 1D line evolves in a different manner to that of the 2D sheet,
this evolution is also quantified by the elements of $\mathbf{F}$,
albeit a different combination to that of the 2D sheet. As such, there
is not a unique scalar stretching rate that characterises fluid
deformation relevant to all transport phenomena but rather an evolving
tensorial quantity, the elements of which are relevant for different
initial and boundary conditions.

While pore-scale imaging provides topographic data for specific porous samples, it is necessary to develop stochastic models of dispersion and mixing from these datasets under appropriate assumptions of stationarity and ergodicity. This approach facilitates the development of upscaled transport models which both honour the pore-scale dynamics and are derived directly from the pore-scale architecture. This approach revolves around development of a CTRW model of fluid deformation, the stochastic kernels of which are populated from pore-scale computational fluid dynamics simulations. This approach is supported by observations over several different 2D and 3D steady flow fields~\cite{Le-Borgne:2008aa,Le-Borgne:2008ab,Lester:2016ab} that fluid advection (and hence deformation) follows a spatial Markov process along fluid streamlines. Given computation of the steady fluid velocity field $\mathbf{v}$ and velociy gradient $\nabla\mathbf{v}$ to sufficient precision over the pore-space, the fluid deformation evolves as per (\ref{eqn:deform3D}).

Recent studies~\cite{Lester:2016aa,Lester:2016ab,Dentz:2015aa} have
shown that by rotation of (\ref{eqn:deform3D}) into a streamline
coordinate system, the fluid deformation evolution equation
(\ref{eqn:deform3D}) takes on a particularly simple form, and permits
explicit solution in terms of a CTRW along fluid streamlines.  As the
elements of the velocity gradient in this frame are almost fully
independent and Gaussian distributed for random
flows~\cite{Lester:2016aa}, stochastic models for the evolution of
$\mathbf{F}(t)$ can be directly determined from pore-scale fluid
dynamics simulations. This stochastic characterisation of tensorial
fluid deformation generates a richer dataset than is given by, for
example, the Lyaponuv exponent or algebraic stretching rate alone, and
so can be used as inputs for lamellar-based models of dilution and
mixing relevant to various transport phenomena.

Under this approach, exponential fluid stretching due to pore-scale
chaos is captured along with shear deformation due to viscous flow,
facilitating characterisation of mixing and dispersion in both the
pre-asymptotic and asymptotic regimes.  The development permits
application of lamellar mixing
models~\cite{Duplat:2010aa,Le-Borgne:2015aa} to predict evolution of
concentration PDF within a solute plume, as well as augmented
longitudinal and transverse dispersion which explicitly capture the
deformation dynamics inherent to the porous architecture. These
stochastic models automatically capture pore-scale chaotic advection,
and facilitate the development of quantitative linkages between
pore-scale geometry (such as topology, tortuously, heterogeneity etc)
and the stochastic kernels.  In the absence of such tools, it is difficult to envisage how to process the
large datasets associated with CFD modelling of pore-scale geometries
beyond direct simulation of transport and mixing in the specific
porous architectures at hand.  Stochastic modelling of fluid
deformation and lamellar mixing addresses this need and thus provides
the building blocks for upscaling methods for mixing and transport
which explicitly honour both the pore-scale deformation dynamics and
architecture.

\section{Conclusions}
There is considerable empirical evidence dating back to the 1960's to
show that the mixing and spreading of solute plumes as they travel
through saturated porous media display unexpected dynamics and
statistics, at both short and long travel times. Early researchers
sought to reconcile these characteristics by identifying upscaled
('effective') quantities at the representative elemental volume (REV)
scale or larger, e.g. macrodispersivities, bulk reaction rates
etc. While consistent with the underlying picture of (laminar) Darcian
flow at the meso- and macroscale, these approaches neglected the
complexity of fluid dynamics at the pore scale in favour of unduly
homogenized descriptions. In recent years, as observation and
computational techniques have advanced, it has become possible to
study transport processes at the pore scale and to embark upon
quantitative upscaling efforts to support estimation of effective
parameters at the macroscale.  Even so there is a richness of physics
operating at the pore scale that can be overlooked in the rush to find
robust upscaling methodologies.

We have demonstrated in this paper that even steady, non-turbulent
flows must engender chaotic (space-filling) fluid trajectories in all
3D random porous media, with an idealized 3D random pore network as an
illustrative example.  This profound result is a consequence of the
topological complexity inherent to all porous media and presents a
fundamentally new genesis for observed microscale plume spreading,
that has previously been categorized as mechanical dispersion
generated by streamline branching, as depicted in
Figure~\ref{fig:2Dstreamlines}.  Indeed, kinematically speaking, the
process of steady streamlines separating around grain boundaries
cannot lead to persistent plume spreading in 2D and, as was shown in
this work, persistent spreading can only be generated by steady
continuum flows in 3D.  Chaotic advection in steady 3D pore-scale flow
generates finely striated material distributions which, when coupled
with molecular diffusion, leads to mixing and dispersion dynamics that
are incongruent with traditional pictures of mechanical dispersion.

The topological approach we have outlined allows fundamental
characteristics of flow and transport to be studied independently of
the vast complexities of pore geometry, surface roughness and chemical
reaction that are so important in digital reconstructions of
pore-fluid interactions.  Idealised pore network models show that pore
fluids deform and mix rapidly through successive pore branch-merge
transitions. The addition of molecular diffusion allows tracer
distributions to be estimated, with fundamentally different asymptotic
plume behaviours exhibited by 2D and 3D flows.  Plume maximum
concentrations and variances decline algebraically with longitudinal
distance in 2D (\ref{eqn:cm2D}), whereas in 3D this decay is
exponential (\ref{eqn:cm3D}).  Chaotic advection inherent in 3D porous
networks also retards temporal scaling of longitudinal dispersion
(\ref{eqn:asymptote}) to be super-diffusive rather than ballistic in
the absence of molecular diffusion, and these chaotic signatures
persist at the macroscale.  These dynamics persist in the presence of
molecular diffusion, where accelerated transverse mixing due to
chaotic advection retards longitudinal dispersion as per
(\ref{eqn:long_dispersion}).  These augmented mixing and dilution
dynamics have significant implications for effective reaction rates at the macroscale.

With these results in mind it is now appropriate to re-visit the
conventional hydrodynamic dispersion conceptualisation depicted in
Figure~\ref{fig:2Dstreamlines}. Molecular diffusion will always be present, as will the mechanical
branching and channelling of flow around sediment grains and into pore
throats.  However the inherently 2D shear-dominated fluid deformation dynamics may now be
seen to be an over-simplification and is potentially misleading in
real systems.  Rather, mechanical dispersion in natural porous media
is best viewed as a consequence of the continued stretching and
folding experienced by genuinely 3D flows within assemblies of
microscale pores and connections. Thus chaotic advection is at the
very heart of mechanical dispersion and it contributes to accelerated
and persistent longitudinal and transverse spreading.  It is vital
that stochastic models capture these dynamics to facilitate the
development of {\em physically consistent\/} upscaled transport
theories.  Both the topology and geometry of the pore space make
interlinked contributions to mechanical dispersion and, as we have
shown, all random pore space topologies and geometries generate
pervasive chaos, even the simple and highly idealised ones considered
in this work.
 
These fundamental topological results are independent of the
geological and chemical complexity that is so prevalent in real
systems.  Nevertheless, topological influences on flow and transport
will always be present in observation and experiment.  An important
question is how to partition the observed net (upscaled) transport
behaviours into contributions from the underlying component processes
and phenomena.  Such a partitioning approach may be assisted by
quantifying topological contributions to upscaled transport
characteristics, and this paper has sought to take an early step down
this path by focusing on the steady, low-energy flow domain.  Transient processes (e.g. time-dependence, reaction
kinetics) and threshold phenomena (e.g. immiscible phase displacement)
await pore-scale topological analysis.  Much more research needs to be
done to develop a comprehensive picture of how the full richness of
pore-scale physics expresses at the REV scale and beyond.

\appendix
\section{Pore Network Model}\label{appendix:pore_model}

To compose a 3D random open porous network model situated within the
semi-infinite domain
$\mathcal{D}:\{x,y,z\}\in(-\infty,\infty)\times(-\infty,\infty)\times[0,\infty)$,
where $z$ is the mean flow direction, we consider a series of
so-called ``mapping planes'' (Figure~\ref{fig:pore_network}(a))
oriented parallel to the $xy$-plane, which are distributed along the
$z$-axis at integer multiples of the dimensionless pore element length
$L$. For each realization of the random network model, the location of
non-overlapping pores of dimensionless uniform diameter $d$ are
prescribed by a random process within each mapping plane, and pore
branch and merge elements are then used to connect pores between
adjacent mapping planes to form a continuous 3D network.

To simplify composition, the network model is constrained to be
periodic in the $(x,y)$ plane, such that the domain $\mathcal{D}$ is
comprised of reflections of the semi-infinite cell
$\mathcal{D}_0:\{x,y,z\}\in[0,1]\times[0,1]\times[0,\infty)$. The
number of pores within each mapping plane in $\mathcal{D}_0$ is $3J$,
for some integer $J>0$, and the location of the $j$-th pore in the
$n$-th mapping plane in $\mathcal{D}_0$ is then labelled
$\mathbf{r}_{n,j}=(x_{n,j},y_{n,j},n L)$. The non-overlapping
condition is then
$||\mathbf{r}_{n,j}-\mathbf{r}_{n,k}+\delta\mathbf{r}||> d\, \forall
j\neq k$
and $\delta\mathbf{r}=(j_1,j_2,0)$ $\forall j_1,j_2\in\{-1,0,1\}$.

To construct the 3D network constrained to these locations, pores
$j=1:J$ within each mapping plane are labelled as ``branch'' pores
(i.e. pores about to branch), and the remaining pores $j=J+1:3J$ are labelled as
``merge pores'' (i.e. about to merge), as per
Figure~\ref{fig:pore_network}(a). A merge pore at plane $n$ is
connected to two branch pores at plane $n+1$ by a pore branch element
located between these planes, and conversely two branch pores at plane
$n$ are connected with a single merge pore at plane $n$ by a pore
merge element.

Connections between merge and branch pores in adjacent planes are made by identifying unique nearest neighbour groupings of a single merge pore in one plane and two branches in the adjacent plane (accounting for periodicity in the $xy$-plane), such that the following sum is minimized
\begin{equation}
\begin{split}
\sum_{j=1}^J \sum_{k=J+1}^{3J} I^b_{n,j,k}||\mathbf{r}_{n,j}-\mathbf{r}_{n+1,k}+\delta\mathbf{r}||_\bot^2+\sum_{j=J+1}^{3J} \sum_{k=1}^{J} I^m_{j,k}||\mathbf{r}_{n,j}-\mathbf{r}_{n+1,k}+\delta\mathbf{r}||_\bot^2,
\end{split}
\end{equation}
where $\bot$ indicates the elements in the $xy$-plane only, and $I^b_{n,j,k}\in\{0,1\}$ ($I^m_{n,j,k}\in\{0,1\}$) is the indicator function that pore $j$ in mapping plane $n$ branches (merges) to connect with pore $k$ in plane $n+1$. Note that groupings are restricted such that a pair of branch pores at plane $i$ common to a single merge pore at $n-1$ do not share the same merge pore at $n+1$ - this restriction eliminates ``degenerate'' pore branch/merger couplings which are isolated from the global network. As all pores undergo sequential branching and merging as they propagate along the $z$-coordinate, the volumetric flow-rate within each pore type is preserved (where merge pores have half that of branch pores) throughout the network. Whilst this topology of sequential pore branching and merging is a small subset of the general topology of random porous networks, this minimum complexity is inherent to all porous media and is sufficient to generate chaotic advection~\cite{Lester:2013aa}.

To accommodate connections between the merge pore locations $\mathbf{r}_{m_1},\mathbf{r}_{m_2}$ and the branch pore location $\mathbf{r}_{b_1}$ at adjacent mapping planes, the pore branch/merge element $\Omega$ is rotated and stretched such that the $\hat{\mathbf{e}}_x$, $\hat{\mathbf{e}}_y$ vectors in Figure~\ref{fig:pore_branch} respectively are aligned along $\mathbf{r}_{m_1}-\mathbf{r}_{m_2}$, $\mathbf{r}_{b_1}-\frac{1}{2}(\mathbf{r}_{m_1}+\mathbf{r}_{m_2})$, and the merge pore center-to-centre distance $L_2$ and $\hat{\mathbf{e}}_3$ extent $L_3$ of $\Omega$ are re-scaled as
\begin{align}
&L_2\mapsto||\mathbf{r}_{m_1}-\mathbf{r}_{m_2}||,\\
&L_3\mapsto||\mathbf{r}_{b_1}-\frac{1}{2}(\mathbf{r}_{m_1}+\mathbf{r}_{m_2})||,
\end{align}
and the orientation angle $\theta$ in the $(x,y)$ plane of a pore branch/merge element $\Omega$ is then
\begin{equation}
\theta=\arctan\left(\frac{(\mathbf{r}_{m_1}-\mathbf{r}_{m_2})\cdot\hat{\mathbf{e}}_y}{(\mathbf{r}_{m_1}-\mathbf{r}_{m_2})\cdot\hat{\mathbf{e}}_x}\right).\label{eqn:twistangle}
\end{equation}
As $\theta$ describes the orientation of 2D manifolds (or minimal flux surfaces) in the fluid bulk, the distribution of $\theta$ governs both chaotic advection and transport dynamics throughout the porous network, whether random or ordered~\cite{Lester:2013aa}.

Following the process described above, the set of pore locations $\mathbf{r}_{n,j}$ for $n=0:\infty$, $j=1:3J$ completely defines a single realisation of the 3D open porous network, and the set of all realisations of the porous network form a statistical ensemble which is ergodic and stationary. If the pore locations $\mathbf{r}_{n,j}$ are randomly distributed uniformly and independently within each mapping plane, then due to the elimination of degenerate pore branch/merger couplings, the resultant distribution of $\theta\in[0,\pi]$ is also uniform and independent both within and across mapping planes. As such, the orientation angle $\theta$ follows a Markov process in space, and so forms an appropriate basis for CTRW modelling of transport and mixing.

\section*{Acknowledgements}
The authors wish to thank Regis Turuban for permission to present his computational results in Figure \ref{fig:3Dmanifolds}.


\end{document}